\documentstyle[11pt, preprint]{aastex}
\begin{document}


\def\beq{\begin{equation}}
\def\eeq{\end{equation}}
\def\a{\alpha}
\def\t{\theta}
\def\r{{\it RHESSI}\,}
\def\ie{{\it i.e.}\,}
\def\eg{{\it e.g.}\,}
\def\ea{et al.\,}
\def\3he{$^3$He\,}
\def\he4{$^4$He\,}
\def\eq{equation\,\,}
\def\eqs{equations\,\,}
\def\cc{{\rm cm}^{-3}}
\def\g{\gamma}
\def\loss{{\rm loss}}
\def\Coul{{\rm Coul}}
\def\sy{{\rm synch}}
\def\scat{{\rm scat}}
\def\esc{{\rm esc}}
\def\eff{{\rm eff}}
\def\tr{{\rm cross}}
\def\ic{{\rm IC}}
\def\br{{\rm brem}}
\def\ac{{\rm ac}}
\def\min{{\rm min}}
\def\max{{\rm max}}
\def\tot{{\rm total}}
\def\rt{{\rm rad tot}}
\def\ln{{\rm ln}}
\def\Clog{\ln\Lambda}
\def\ltsim{\hbox{\rlap{\raise.5ex\hbox{$<$}}{\lower.7ex\hbox{$\sim$}}}}
\def\gtsim{\hbox{\rlap{\raise.5ex\hbox{$>$}}{\lower.7ex\hbox{$\sim$}}}}

\title{Particle Acceleration in Solar Flares and Enrichment of $^3$He and Heavy
Ions}

\author{Vah\'{e} Petrosian\altaffilmark{1,2,3}}

\altaffiltext{1}{Department of Physics, Stanford University, Stanford, CA, 94305
email; vahep@stanford.edu}
\altaffiltext{2}{Kavli Institute for Particle Astrophysics and Cosmology,
Stanford University, Stanford, CA 94305}
\altaffiltext{3}{Also Department of Applied Physics}

\begin{abstract}
We discuss possible mechanisms of acceleration of particles in solar flares and
show that turbulence plays  an 
important role in all the mechanism. It is also argued that  stochastic particle
acceleration by turbulent plasma waves is the most likely mechanism for
production of the high energy electrons and ions responsible for observed
radiative signatures of solar flares and for solar energetic particle or SEPs,
and that the predictions of this model agrees well with many past and recent
high spectral and temporal observations of solar flares. It is shown that, in
addition, the model explains many features of   SEPs  that accompany flares. In
particular we show that it can  successfully explain the observed extreme
enhancement, relative to photospheric values,  of $^3$He ions and the relative
spectra of $^3$He and $^4$He. It has also the potential of explaining the
relative abundances of most ions including the increasing  enhancements of heavy
ions with ion mass or mass-to-charge ratio.

\end{abstract}

\section{Introduction}

The aim of this paper is to investigate to what extent current theoretical
models of
acceleration of particles can explain the observed characteristics of solar
energetic particles (SEPs), in particular the extreme enhancement,
relative to photospheric values, of $^3$He ions in many flares, in particular in the so-called
impulsive (or He-rich) events. In the next section we  present a general
description
of the possible acceleration mechanisms for production of particles in solar
flare.  We show that plasma waves or turbulence play a major
role in all these processes. In \S 3 we describe the model of
stochastic
acceleration by plasma turbulence in some detail and review its successes in
describing observations of solar flares, in particular the radiative signatures
produced by accelerated electrons and protons. In \S 4 we address the
problem of observed enhancements
of \3he and heavy elements.  Here we first
present a brief review of some of the earlier models proposed for production of
these enhancements and then compare predictions of the stochastic acceleration
model with the observations. A brief summary and discussion is presented in \S
5.

\section{Particle Acceleration}
\label{acceleration}

In this section we first compare various acceleration processes and stress the
importance of plasma wave turbulence  ({\bf PWT}) as an agent of
acceleration, and then describe the basic scenario and equations for treatment
of
these processes in the stochastic acceleration ({\bf SA}) model.
As described below there is growing evidence that PWT plays an
important role in acceleration of particles in general, and in solar flares in
particular. The two most commonly used acceleration mechanisms are the
following.

{\bf 1. Electric Field Acceleration:}  Electric fields parallel to
magnetic fields can accelerate charged particles.
For solar flare conditions the Dreicer field ${\bf E}_D = k_BT/(ev\tau_{\rm
Coul})\sim 10^{-5}$ V/cm. (Here $k_B, T, v$ and $e$ are the Boltzmann constant,
plasma temperature, particle velocity and charge,
and $\tau_{\rm Coul}$ is the mean collision time.)
Sub-Dreicer fields (${\bf E}\leq{\bf E}_D$) can only accelerate particles up to
10's of keV (for a typical flare
length scale
$L\sim10^9$~cm)  which is far below the  10's of MeV electrons or $>10$ GeV
protons
required
by observations%
\footnote{Dreicer field will be large if the
resistivity and density are
anomalously high (Tsuneta 1985; Holman 1996a, b).}.  Super-Dreicer fields,
which seem to be present in many simulations of reconnection (Drake, 2006, see
also
Cassak \ea 2006; Hoshino 2006, see also Zenitani \& Hoshino, 2005),
accelerate particles at a rate that is faster than $\tau_{\rm Coul}$. This can
lead to a runaway and an unstable electron distribution which, as shown
theoretically,
by laboratory experiments and by the above mentioned simulations,
most probably will give rise to PWT (Boris et al.
1970, Holman 1985)%
\footnote{Note also that  production of a broad
power-law requires a wide range of  potential drops
(Litvinenko, 2003).}. {\it In summary the electric fields arising as a result of
reconnection cannot be the sole agent of acceleration, but may produce an
unstable particle momentum distribution which will produce PWT that can then
accelerate particles.}

{\bf 2. Fermi Acceleration:} Nowadays this process has been divided into two
kinds.
In the original Fermi process particles of velocity $v$ scattering
randomly with a (pitch angle cosine $\mu$) diffusion rate $D_{\mu\mu}$ by moving
 scattering agents with a velocity $u$
gain energy at a rate proportional to $(u/v)^2D_{\mu\mu}$.
This, known as a {\it second order Fermi process}, is what we call {\it
stochastic acceleration}
({\bf SA}). {\it For solar flares the most likely agent for scattering is PWT.}
An alternative process is what is commonly referred to as {\bf Shock}
acceleration.
Because particles
crossing the shock (or any kind of flow convergence) gain energy at a rate which
is
$\propto u_{sh}/v$, this is known as the {\it first order Fermi} process.
Ever since the demonstration by several authors that a very simple version of
this process leads to a power law spectrum that agrees approximately with
observations of the cosmic rays, shock acceleration is commonly invoked in space
and astrophysical
plasmas. However, this simple model, though very elegant, has many shortcomings
Specially
when applied to non-thermal radiation processes.

{\bf Shock} acceleration, as commonly used, requires injection of (sometimes
high energy)
particles and is unable to accelerate low energy background particles.
The simple results also break down when  one considers
more realistic models. For example, inclusion of losses
(Coulomb at low energies and synchrotron at high) or influence of
accelerated particles on the shock structure (see \eg Ellison \ea, 2005: Amato
\& Blasi 2005)
cause breaks and generally deviations from a simple power law.
More importantly, a {\it shock by itself cannot accelerate particles}
and requires some
scattering agents (most likely PWT) to cause repeated passages of the particles
through the shock front
(thus the name Fermi).  The rate of
energy gain is governed by the scattering rate $D_{\mu\mu}$.
This is the case for the so-called quasi-parallel shock, but  shocks
are most likely quasi-perpendicular in which case, as pointed out by Jokipii
(\eg 1987),
the particles may drift along the surface of the shock and get accelerated more
efficiently. However, this model also suffers from the above mentioned
fundamental limitation and
requires injection of particles with velocities $v>u_{sh}\eta$,
where $\eta \gg 1$ is the ratio of parallel to perpendicular (to magnetic field)
diffusion coefficients. In a more recent simulation of this process Giacolone
(2005)
shows that a non-thermal tail can be produced from a thermal distribution.
However, this model includes
significant field fluctuations, which as advocated below can accelerate low
energy
particles more efficiently
than shocks. Thus, it is not obvious what are the
relative roles of shock vs SA here.
Moreover, for solar flares there is no
direct evidence for shocks near the LT
during the impulsive phase, and some of the features that make
acceleration of cosmic rays by shocks attractive are not present%
{\footnote{For example, acceleration by sub Alfv\'enic shocks is
questionable (see Kulsrud 2005), and the streaming instability,
which is a part of the traditional shock acceleration
can be partially suppressed in the presence of PWT (Yan \& Lazarian 2002;
Farmer \& Goldreich 2004).}.

{\bf Stochastic Acceleration} is favored because the PWT needed for scattering
can also accelerate particles stochastically with the rate $D_{EE}/E^2$,
so that shocks may not be always necessary. First, contrary to common belief,
Hamilton \& Petrosian 1992 and 
Miller \& Reames 1996 have shown that for flare conditions PWT can
accelerate the background particles to high energies within the desired
time.  More importantly, at low
energies or in strongly magnetized plasmas the acceleration rate $D_{EE}/E^2$
may exceed the scattering rate $D_{\mu\mu}$ (see Pryadko \& Petrosian, {\bf
PP97}).  {\it Thus,
for flare conditions low energy particles are accelerated more efficiently by
PWT than
by shocks}. We note, however, that in gradual flares there may be a
need for a second stage acceleration of SEPs by high coronal shocks%
\footnote{In practice, \ie mathematically, there
is little difference between the two mechanisms (Jones 1994), and the
acceleration by turbulence and  shocks can be combined (see below).}.

{\it Irrespective of which process dominates the particle acceleration, it is
clear
that PWT has a
role in all of them. Thus, understanding of the production of PWT and
its interaction with particles  is extremely important.}
Moreover, {\bf turbulence} is expected to be present in most astrophysical
plasmas
including solar flares and in and around interplanetary shocks, because the
ordinary and magnetic Reynolds numbers for these situations are
very large and may be the most efficient channel of energy dissipations in
non-equilibrium systems such as solar flares.
In recent years there has been a substantial progress
in the understanding of
MHD turbulence (Goldreich \& Sridhar, 1995, 1997; Lithwick \& Goldreich 2001;
Cho \& Lazarian 2002 and 2006). These advances provide new tools for a more
quantitative
investigation of turbulence and the role it plays
in solar flares.

\section{Stochastic Acceleration and Solar Flares}
\label{SAandFlares}

\subsection{Basic Scenario}
\label{scenario}

The complete picture of a solar flare involves many phases or steps.
After a complex pre-flare buildup, the first
phase is the reconnection or the energy release process.  The final consequences
of
this released energy are the observed radiations (from long wavelength radio to
GeV gamma-rays), SEPs and CMEs.  Many processes are involved in
conversion of the released energy into radiation. As stressed above
we believe that PWT plays an important role in these processes.  We envision the
following
scenario.
Magnetic energy is converted into turbulence by the reconnection process above
corona
loops which we refer to as the acceleration site or the loop top ({\bf LT})
source (see Figure \ref{model}). The turbulence or waves generated on some
macroscopic scale (some fraction of reconnection site scale) undergo
two kind of interactions. The  first is {\it nonlinear wave-wave interaction}
causing
them to undergo dissipationless cascade to smaller scales. The second  is
{\it wave-particle interaction} which becomes more  important at smaller scales.
This
damps the turbulence. The lost energy goes into heating the background plasma
and/or accelerating particles into a non-thermal
tail.
The accelerated particles on the open field lines escape the Sun and are
observed as SEPs having undergone varied degree of scatterings and acceleration
(\eg by interplanetary shocks) during their transport to the Earth. The escaping
electrons may also produce type III radio bursts. The particles on the field
lines associated with the closed loops spiral down towards the photosphere
and produce the observed
non-thermal radiations
via their interactions with the
background particles (and fields) along the loop and primarily at the footpoints
({\bf FPs}).
However, most of the energy of the non-thermal particles is lost  by collisions
causing heating and evaporation of the colder chromospheric plasma, which is
responsible
for most of the softer thermal radiation. This process,
described by the hydrodynamic equations, has a
time-scale comparable to the sound travel time, and
is somewhat decoupled from the acceleration-transport process that has
a much shorter time-scale. However, the evaporation can modulate the high
energy processes by changing the density and temperature in the
acceleration site.

\subsection{Formalism}
\label{equations}

The mathematical treatment of the processes involved in such a scenario is long
and complex. Below we give an outline of the important aspects.

\subsubsection{Kinetic Equations}
\label{kinetic}

{\bf The spectrum of PWT} $W({\bf k},t)$ is determined by the wave-wave and
wave-particle interactions. The cascade is evaluated  from the rates of
wave-wave interactions. For
example the three wave interactions can be presented as (see \eg Chandran 2005;
Luo \& Melrose 2006)
\beq\label{wave-wave}
\omega({\bf k}_1) + \omega({\bf k}_2) = \omega({\bf k}_3) \,\,\,\, {\rm and}
\,\,\,\, {\bf
k}_1+{\bf k}_2={\bf k}_3,
\eeq
where $\omega$ and ${\bf k}$ are the wave frequency and wavevector. The
interaction
rates can be represented by the wave diffusion coefficient $D_{ij}$. In
general, at large scales the  cascade
time $\tau_{\rm cas}\sim k^2/D_{ij}$ is shorter than the damping time
$\Gamma(k)^{-1}$ resulting from wave-particle interactions. Thus, the waves
undergo dissipationless cascade from the injection scale $k_\min\sim L^{-1}$
(with a rate ${\dot Q}^W({\bf k})$) till $k_\max$ where $\tau_{\rm
cas}\Gamma=1$, known as the {\it inertial range}. Beyond $k_\max$ the
spectrum of waves drops off more rapidly.
Adopting the
diffusion approximation (see \eg Zhou \& Matthaeus 1990), one
can obtain
the evolution of the spatially integrated wave spectrum $W({\bf k}, t)$
\beq
\\{\partial {W} \over \partial t}
= {\partial \over\partial k_i}\left[D_{ij}{\partial\over \partial
k_j}{W}\right] -
\Gamma({\mathbf k}){W} - {{W}\over T^{W}_{\rm esc}({\mathbf k})} + \dot{Q}^{W}.
\label{waves}
\eeq
We will assume that all the wave energy is absorbed so that there is no wave
escape, \ie $T^{W}_{\rm esc}\rightarrow\infty$.

The general equation for treatment of the {\bf particle acceleration and
transport} is the
Fokker-Planck equation
for the gyro-phase averaged particle distribution $f(t,s,E,\mu)$ as a function
of
distance $s$ along the magnetic field lines. This equation is simplified
considerably if one can adopt the
isotropic
approximation (or deal with the pitch-angle averaged distribution) and 
impose
the homogeneity condition (or deal with distribution integrated over the
acceleration region). These are reasonable assumptions for particles at the
flare site and  amount to determination of the evolution of the
energy spectrum
$N(E,t)=\int \int d\mu ds f(t,s,E,\mu)$:
\beq
{\partial N \over \partial t}
 =  {\partial \over \partial E}\left[D_{EE}{\partial N \over \partial E}
 - (A-\dot E_L) N\right]
 - {N \over T^p_{\rm esc}} +{\dot Q}^p,
\label{kineq}
\eeq
Here  $D_{EE}/E^2, A(E)/E$ describe  the energy
diffusion and direct acceleration rates. They are obtained from consideration of
the wave-particle interactions, which are often dominated by resonant
interactions, specially for low beta (magnetically dominated) plasma, such that
\beq\label{resonance}
\omega({\bf k})-k\cos\t v\mu = n\Omega/\gamma,
\eeq
for  waves propagating at an angle $\t$ with respect to the large scale magnetic
field,
and a particle of velocity $v$, Lorentz factor $\gamma$, pitch angle cosine
$\mu$ and gyrofrequency $\Omega$. Here the harmonic number  $n$ (not to be
confused with the density)
is equal to zero for transit
time damping ({\bf TTD}) process. For parallel propagating waves ({\bf PPWs})
$n=\pm 1$
but for obliquely propagating waves
$n= \pm 1, \pm 2, ...$\,.
${\dot E}_L/E$ is the energy loss rate of the particles (due mainly to Coulomb
collisions and synchrotron losses) and $\dot{Q}^p$ and the terms with the
escape times $T_{\rm esc}$ describe the source and leakage of particles.  (For a
more detailed discussion leading to the above equations see Miller et al. 1996
and Petrosian \& Liu 2004, {\bf PL04}).

The above two kinetic equations are coupled by the fact that
the coefficients of one depend on the spectral distribution of the other.
Conservation of energy requires that the energy lost by the waves ${\dot
{\cal
W}_{\rm tot}}\equiv \int \Gamma({\bf k})W({\bf k})d^3k$ be equal to the energy
gained by the
particles from the waves;  ${\dot {\cal E}}=\int [A(E)-A_{\rm sh}]N(E)d E$.
Representing the
energy transfer rate between the waves and particles by
$\Sigma ({\bf k}, E)$ this equality implies that
\beq
\label{coefficients}
\Gamma({\bf k})= \int_0^\infty \d E N(E)\Sigma({\bf k}, E),\,\,\,\,
A(E)=\int_0^\infty d^3kW({\bf k})\Sigma ({\bf k}, E)+ A_{\rm sh}\ ,
\eeq
where we have added $A_{\rm sh}$ to represent
contributions of other (non-SA) processes affecting the direct acceleration, \eg
shocks.

The transport effects and further acceleration of the particles as they travel
in the interplanetary space can also be treated by similar equations.

\subsubsection{Dispersion Relations}

At the core of the evaluation of all the coefficients of the kinetic equations
described above lies the
plasma dispersion relation $\omega({\bf k})$ which determines the
characteristics of the waves that can be excited in the plasma, and the rates of
wave-wave and wave-particle interactions.
For example, in the MHD regime for a cold plasma
\beq
\label{mhddisp}
\omega=v_{\rm A}k\cos\t\,\,\,\,\,  {\rm and}\,\,\,\,\,  \omega=v_{\rm A}k
\eeq
for the Alfv\'en  and fast magneto-sonic waves, respectively, where $v_{\rm A}$
is the Alfv\'en velocity. Beyond the MHD regime a multiplicity of wave modes can
be
present and the dispersion relation is more complex and is obtained from the
following 
expressions:
\beq\label{dispgeneral}
\tan^2\theta = {-P(n_r^2-R)(n_r^2-L)\over (Sn_r^2-REL)(n_r^2-P)}
\eeq
where $n_r=kc/\omega$ is the refractive index, $S={1\over 2}(R+L)$, and
\beq\label{dispterms}
P= 1-\sum_i{\omega_{pi}^2\over \omega^2},\ \ \  R=1-\sum_i{\omega_{pi}^2\over
\omega^2}\left({\omega\over 
\omega+\epsilon_i\Omega_i}\right),\ \ \ {\rm and} \ \ \
L= 1-\sum_i{\omega_{pi}^2\over \omega^2}\left({\omega\over 
\omega-\epsilon_i\Omega_i}\right).
\eeq
Here $\omega_{pi}^2={4\pi n_i q_i^2\over m_i}$ and
$\Omega_i={|q_i|B\over m_ic}$ are the plasma and gyro frequencies,
$\epsilon_i={q_i\over |q_i|}$, and $n_i$, $q_i$, and $m_i$ are the  
density, charge, and mass of the background particles (electron, proton and
$\alpha$.)
Figure \ref{dispersion} shows the more complete picture of the dispersion
surfaces  (depicted by the
curves) obtained from the well known cold plasma expressions
along with the resonant planes
in  the ($\omega,\ k_\parallel,\ k_\perp$) space. Intersections
between the dispersion
surfaces and the resonant planes define the resonant wave-particle interactions.

\begin{figure}[hbtp]
\centerline{
\includegraphics[height=8.5cm,angle=270]{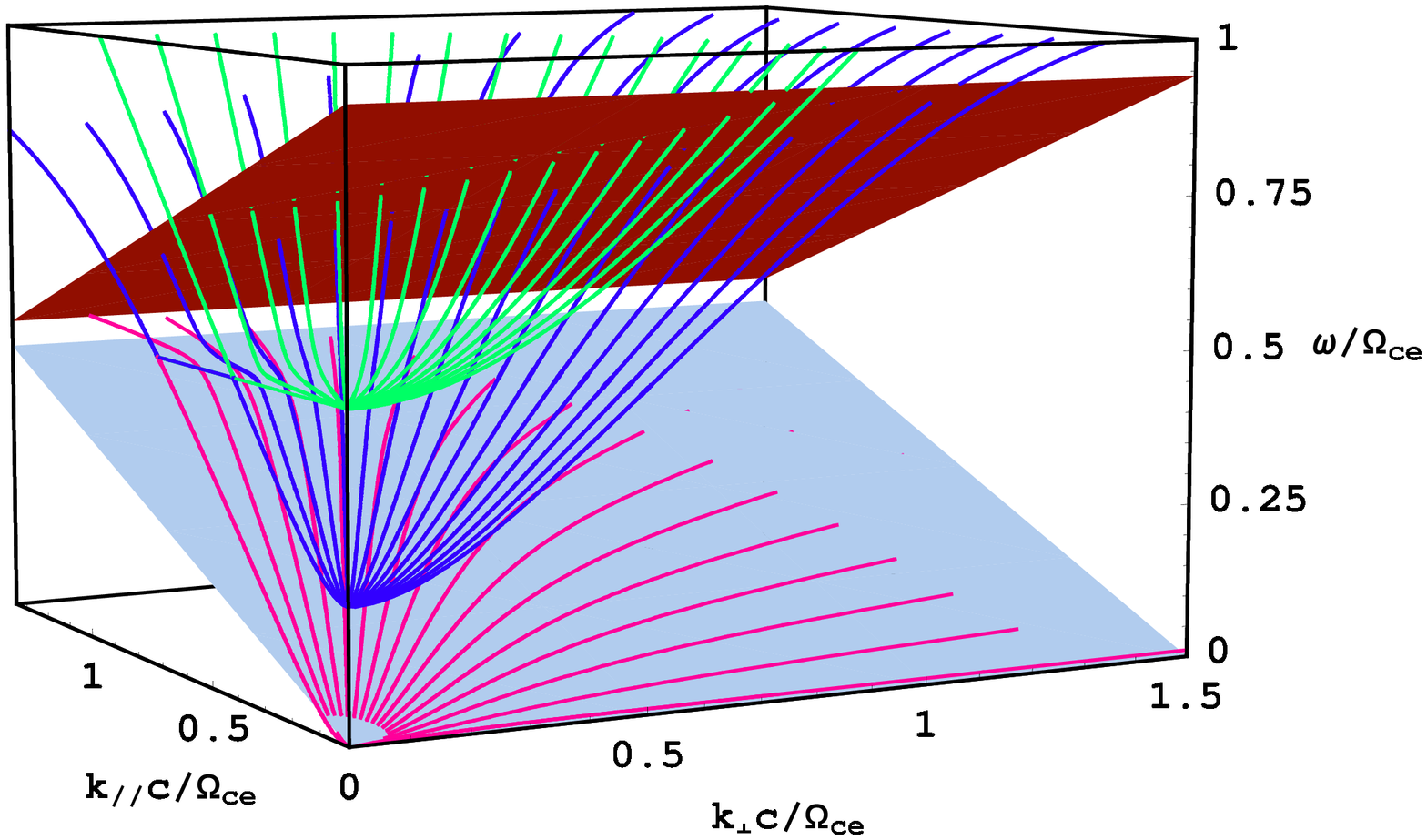}
\includegraphics[height=8.0cm,angle=270]{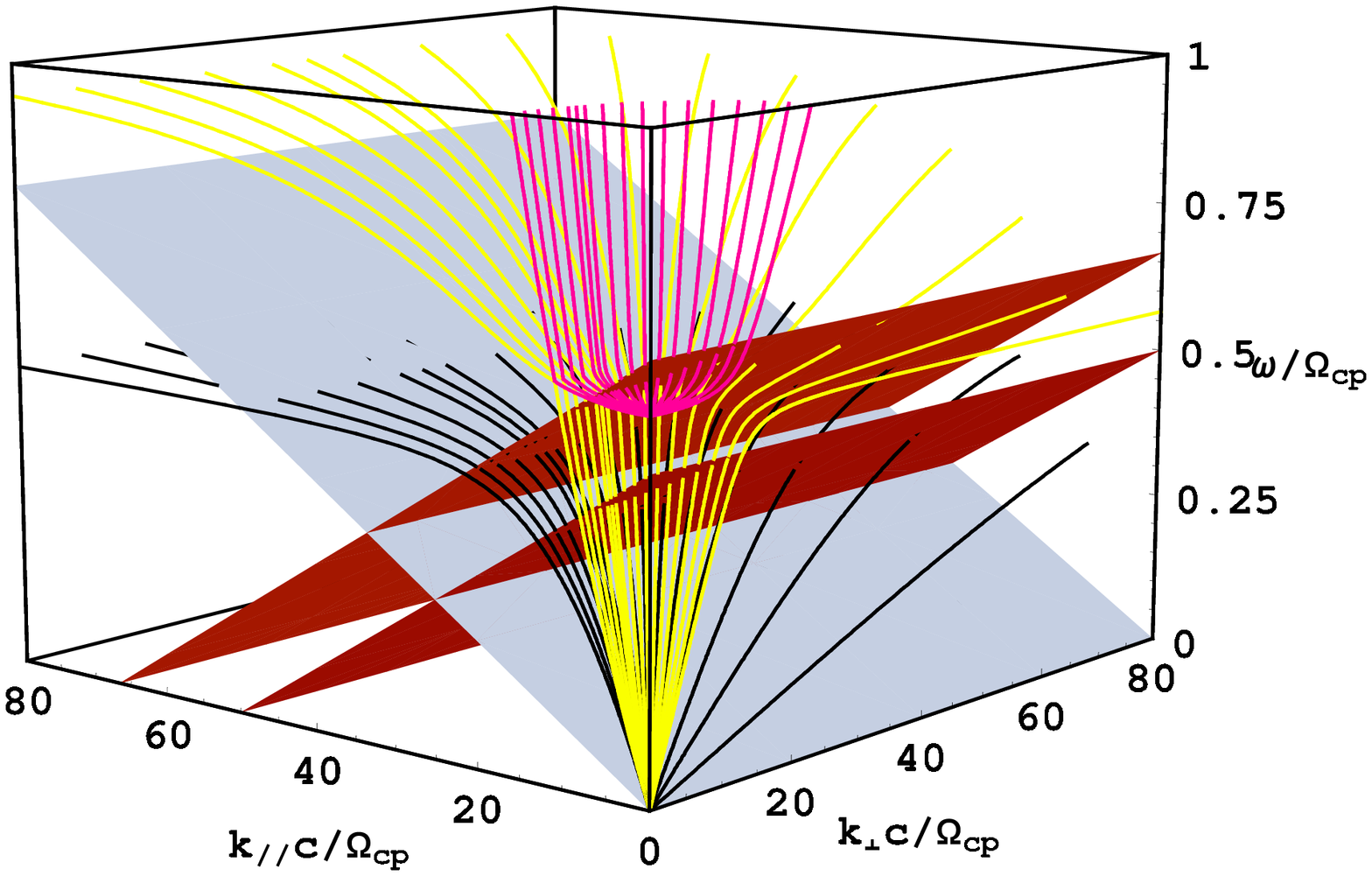}
}
\caption
{\scriptsize
Dispersion relation (curves) and resonance condition (flat) surfaces
for a cold fully ionized H and He (10\% by number) plasma with
$\beta_{\rm A}=v_{\rm A}/c=0.012$
showing the regions
around the electron (left) and proton (right) gyro-frequencies. Only
waves with positive $k_\|, k_\perp$ (or $0<\t<\pi/2$) are shown.
The mirror image with respect to the ($\omega,\ k_{\perp}$) plane
gives the waves propagating in the opposite direction. From high to
low frequencies, we have one of the electromagnetic branches (green),
upper-hybrid
branch (purple), lower-hybrid branch, which also includes the whistler waves
(pink),
fast-wave branches (yellow), and Alfv\'{e}n
branch (black). The effects of a finite temperature modify these curves at
frequencies $\omega\sim kv_{\rm th}$, where $v_{\rm th}=\sqrt{2k_{\rm B}T/m}$ is
the thermal velocity (see \eg Andr\'{e} 1985). The
resonance surfaces are for electrons with $v=0.3c$ and $|\mu|=1.0$ (left: upper
$n=1$, lower $n=0$) and
$^4$He (right: middle $n=1$) and $^3$He (right: upper $n=1$) ions with
$|\mu|=1.0$ and $v=0.01c$. The
resonance surfaces for the latter two are the same when $n=0$ (right: lower).}
\label{dispersion}
\end{figure}

The above dispersion relations are good approximations for low beta plasmas,
\beq
\label{beta}
\beta_p=2(v_s/v_{\rm A})^2=8\pi nk_{\rm B}T/B^2=3.4\times 10^{-2}(n/10^{10}
{\rm cm}^{-3})(100 {\rm G}/B)^2(T/10^7{\rm K})\ll 1,
\eeq
where $v_s=\sqrt{k_{\rm B}T/m_p}$ is the sound speed. For higher beta
plasmas,
\eg  at higher temperatures, these relations
are modified, specially for higher frequencies $\omega\sim kv_{\rm th}$, where
$v_{\rm th}=\sqrt{2k_{\rm B}T/m}$. For example, in the MHD regime, in
addition
to the Alfv\'en mode one gets fast and slow modes with the  dispersion relation
(see \eg Sturrock 1994)
\beq
\label{mhddisp2}
(\omega/k)^2={1\over 2}\left[(v_{\rm A}^2+v_s^2)\pm \sqrt{v_{\rm
A}^4+v_s^4-2v_{\rm A}^2v_s^2\cos {2\t}}\right],
\eeq
and the more general dispersion relation is modified in a more complicated way
(see \eg Andr\'{e} 1985 or Swanson 1989). The finite temperature imparts an
imaginary part to the wave frequency that gives the (Landau) damping rate of the
waves (see \eg Swanson 1989 or Pryadko \& Petrosian 1998, 1999 and Cranmer \&
van Ballegooijen 2003 for application
to solar flare conditions). In general, these rates and the
modification of the dispersion relation are known for Maxwellian (sometimes
anisotropic) energy distribution of the plasma particles. For non-thermal
distributions the damping rates can be evaluated as described in Petrosian, Yan
\& Lazarian (2006) using the
coupling described in equation  \ref{coefficients}. If the damping due to
non-thermal particles is important then the wave and particle kinetic equations
(\ref{waves}) and (\ref{kineq}) are coupled. This will be important in large
intense flares where a substantial fraction of particles are accelerated into a
non-thermal
tail. However, most often the damping rate is
dominated by the background thermal particles so that the wave and non-thermal
particle kinetic
equations decouple. This simplifies matters considerably. One can then use the
imaginary part of the wave frequency for evaluation of the damping rate using
the well-known hot plasma dispersion relation. Moreover, the 'thermal' effects
change the real part of the wave frequency only slightly so that
often the real part (and the particle diffusion coefficients) can be evaluated
using the simpler cold plasma dispersion relation depicted in Figure
\ref{dispersion}.

\subsubsection{Some Model Results} 
 
We now demonstrate some of the basic characteristics
of our scenario which can be compared with observations. The complete treatment
of the problem requires solution of the coupled wave particle kinetic equations
for particles at all energies and pitch angles and waves propagating at all
angles. In principle, given the plasma density, magnetic field, temperature
($n,\
B,\ T$), the geometry of the
region
(represented by a size $L$ here), the rate and scale of injection  of
turbulence $\dot{Q}^W({\bf k}, t)$, one can evaluate the coefficients of
equations
(\ref{waves}) and (\ref{kineq}) and solve the coupled kinetic equations for
determination
of the resultant distributions $N(E,t)$ and $W({\bf k},t)$. However,
under certain circumstances some simplifications are possible.
As mentioned above in most cases the damping by non-thermal particles can be
ignored and the equations are decoupled. The wave spectrum then can be
approximated by a power law in the inertial range ($W\propto k^{-q}, k_{\rm
min}<k<k_{\rm max}$) with a steeper falling spectrum beyond this range. The
exponent $q$ will be in the range expected in a Kolmogorov or Kraichnan cascade.
Furthermore, as shown in Petrosian \ea (2006), the upper limit $k_{\rm max}$
depends strongly on the angle of propagation $\t$ with only quasi-parallel
propagating waves {\bf PPWs} reaching small enough scale to be in resonant with
low energy
particles. Thus, the initial acceleration of thermal particles will be dominated
by such waves. Consequently, in what follows here and the next section we assume
PPWs with the above spectrum. The following
results are based on PL04, where  more details can be found. 

\begin{figure}[htbp]
\leavevmode
\centerline{
\includegraphics[width=7.5cm]{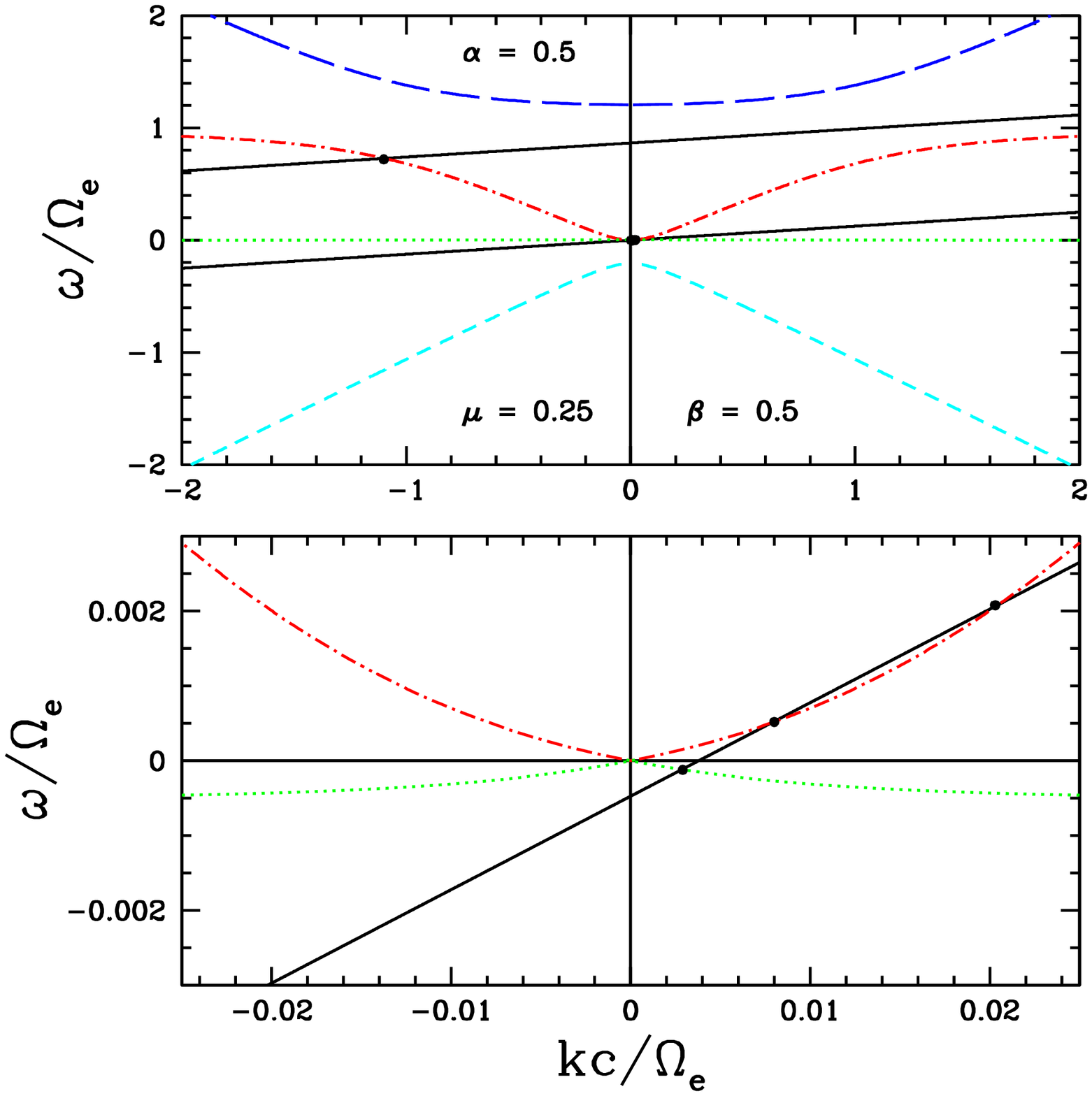}
\includegraphics[width=7.5cm]{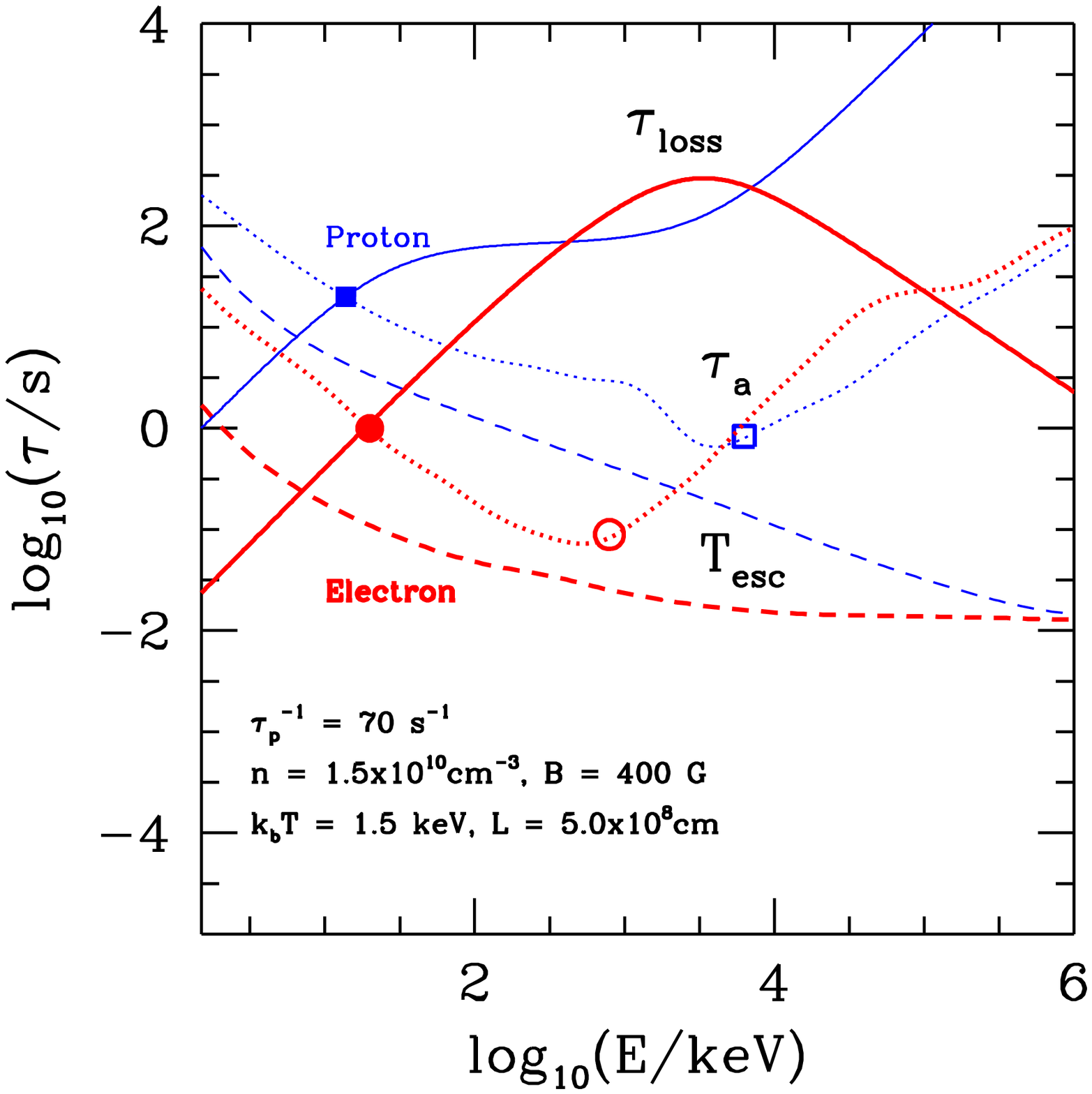}
}
\caption{\scriptsize
{\bf Left panel:} The dispersion relation for PPWs (curves)  and the resonant
relation (lines, from eq.[\ref{resonance}]) for electrons (top) and protons
(bottom, expanded near the origin). Note that the resonant lines start (at
$k=0$) from the gyro frequency of the particle (divided by the Lorentz factor
$\gamma\sim 1$).
{\bf Right panel:} Time scales for a SA acceleration model for solar flare
conditions and an
assumed
spectrum of PWT. Model parameters are indicated in the figure
($\tau_p^{-1}\propto \Omega_e{\cal W}/B^2$ is a
characteristic wave-particle interaction rate and $\alpha\propto \sqrt{n}/B$).
The acceleration, escape, and
loss times are indicated by the dotted, dashed, and solid curves, respectively.
The thick
(red) curves are for electrons and thin (blue) curves are for protons.}
\label{timescales}
\vspace{-0.5cm}
\end{figure}

The left panel of Figure \ref{timescales} shows the
dispersion relation for PPW and the resonant conditions (straight lines) for an
electron and a proton of the specified energy and pitch angle cosine. The main
parameter determining
these curves and the resulting resonances (intersections of the lines with
the curves) is the ratio of the plasma to gyro frequencies
\beq\label{alpha}
\alpha=\omega_{pe}/\Omega_e=(m_e/m_p)^{1/2}/\beta_{\rm A}=3.2(n/10^{10}
\cc)^{1/2}(100 {\rm G}/B).
\eeq
The right panel of Figure \ref{timescales} shows the timescales associated with
the terms in equation (\ref{kineq}), {\it i.e.} loss
time
$\tau_{\rm loss}=E/{\dot {E}_L}$, diffusion time $\tau_{\rm diff} \sim
E^2/D_{EE}$, acceleration time $\tau_{\rm a}=E/A(E)$, and escape time%
\footnote{This is an approximate
relation valid for scattering time $\tau_{\rm scat}\ltsim D_{\mu\mu}^{-1}>$
longer and
shorter than $T_{\rm cross}\sim L/v$, the time for particles to cross the
turbulent region
of size $L$ (see PL04).}$T_{\rm esc}= T_{\rm cross}(1+T_{\rm cross}/\tau_{\rm
scat})$.

\begin{figure}[htbp]
\leavevmode
\centerline{
\includegraphics[width=7.5cm]{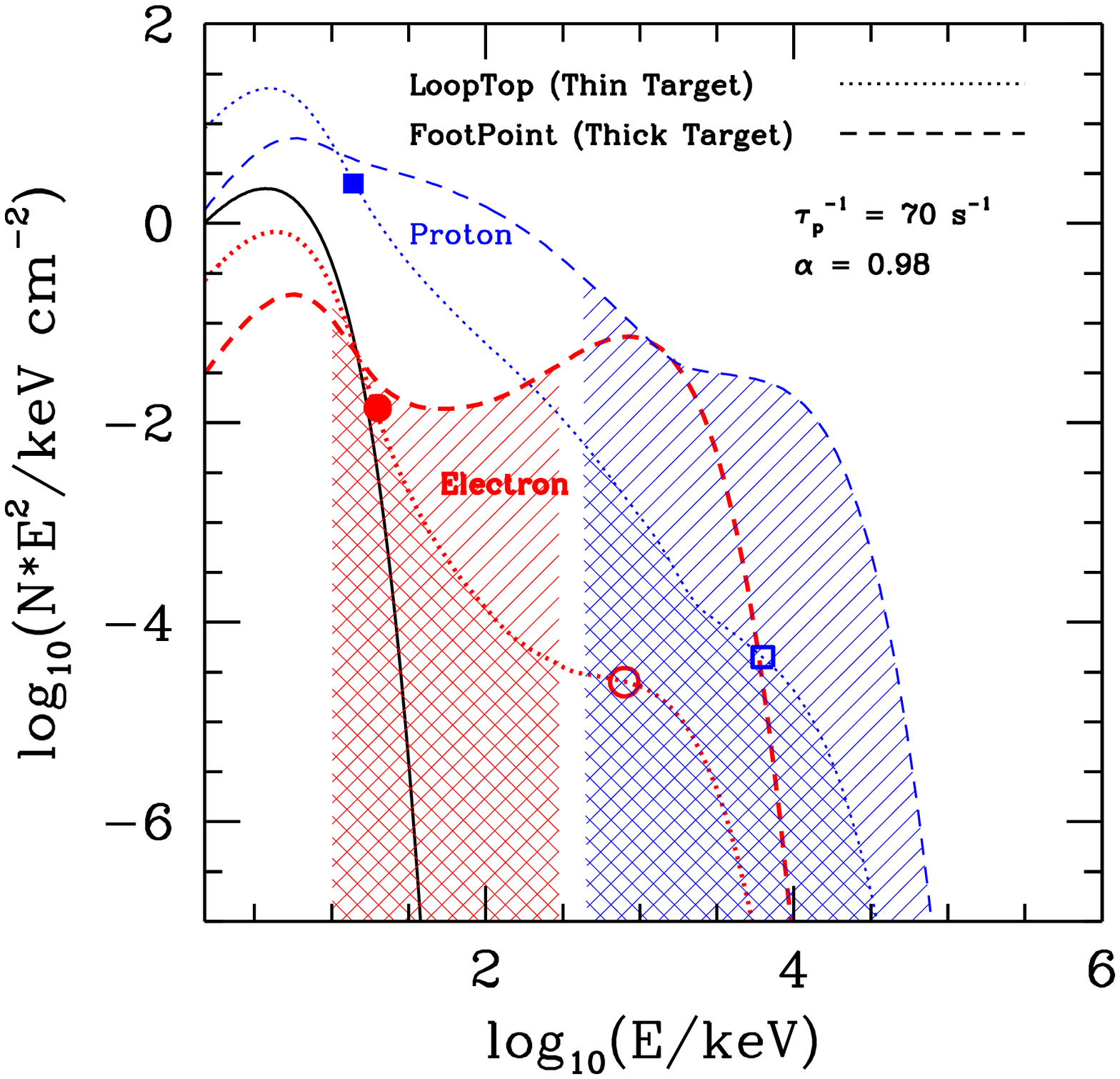}
\includegraphics[width=7.5cm]{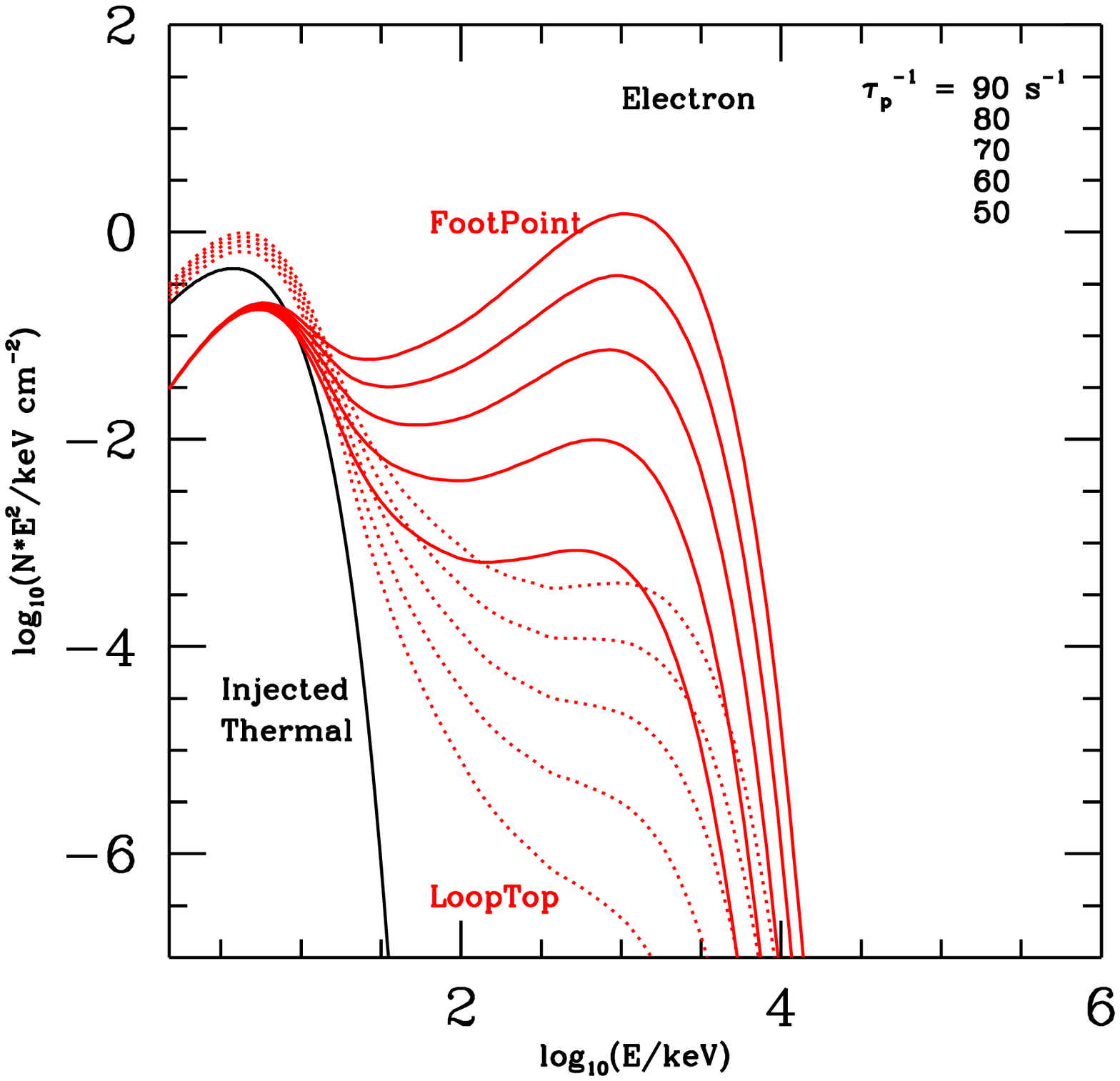}
}
\caption{\scriptsize
{\bf Left panel:}
The spectra of accelerated electrons and protons corresponding to the timescales
shown in Figure \ref{timescales} (right). The dotted
curves give the spectra at the acceleration site (LT) and the dashed curves give
the thick target equivalent
spectra of escaping particles. The solid (black) line
gives the shape of the background (injected) particle distribution. The hatched
areas show the energy ranges for production of  frequently observed ranges of
HXRs (for electrons) and gamma-ray lines (for protons). Spectral features occur
at
the intersections and divergences of these timescales (circle and square signs
shown here and in Fig. \ref{timescales}; right).
{\bf Right panel:} The dependence of electron spectrum on the  ${\cal
W}\propto\tau_p^{-1}$.
Higher ${\cal W}$'s produce harder spectra
and more acceleration vs heating (see PL04 ).
}
\label{spectra}
\vspace{-0.5cm}
\end{figure}

{\it The spectral characteristics} of the particles
arise from the interplay among the
timescales shown above.
Figure \ref{spectra} (left) shows
the corresponding spectra  of electrons and protons at the LT
($N_{\rm LT}$, dotted) and the equivalent FP thick target spectra (dashed lines;
see \eg Petrosian 1973, Park, Petrosian \& Schwartz  1997)
\beq\label{thick}
N_{\rm FP}(E)={\dot E}^{-1}_L\int_0^E N_{\rm LT}(E')T_{\rm esc}^{-1}(E')dE'.
\eeq
Spectral breaks
occur at energies (circles and squares) when different terms
become important. As evident, in general, SA by PWT  produces a
``quasi-thermal''
and a harder non-thermal component; \ie it both ``heats" the plasma and
accelerates
particles. As also seen in the right panel of Figure \ref{spectra}, the relative
strength of the two components depends
sensitively on the intensity ${\cal W}\propto \tau_p^{-1}$ of the PWT. For
higher densities of turbulence the spectra are harder. In addition,
the LT spectra are dominated by the
thermal component with a steep non-thermal component, while the FP  spectra
are harder with little or non-thermal part.  These agree with the observed
features discussed below.
For low ${\cal W}$, \eg during the build-up
and decay phases of a flare, one expects only plasma heating with essentially no
non-thermal component (see below and PL04 ).

\subsection{Application to Solar Flares}
\label{observe}

Ultimately observations must determine the validity of the models.
As described below there have been  extensive comparisons of the SA model with
observations. To our knowledge, there have not been similarly detailed
comparisons for electric field or shock acceleration.

\subsubsection{Radiative Signatures of Electrons}
\label{electron}

{\bf Spatial Structure and Evolution:}
The most direct evidence for the presence of
PWT is observations by {\it Yohkoh} of distinct LT impulsive HXR emission
(Masuda et al. 1994, Masuda 1994), which  have been
shown to be common to almost all {\it
Yohkoh} (Petrosian, et al. 2002) and {\it RHESSI} (Liu et al. 2004) flares.
Petrosian \&
Donaghy (1999) have also shown that these observations require an enhanced pitch
angle scattering to confine the particles near the LT. Coulomb collisions cannot
be
this agent because of high losses they entail.
The most likely scattering agent is PWT, which can also accelerate
particles. Figure \ref{model} (left) shows a  cartoon of the
reconnection and acceleration site
for a flaring loop with the red foam representing the PWT produced
during reconnection. As reconnection proceeds larger closed loops
are formed below this site. This simple picture then predicts  a gradual rise
of the LT
source accompanied with a continuous increase in separation of the FPs. Most
strong flares are
complex and involve multiple (or arcade) of loops (\eg July 23, 2002 flare;
Krucker \ea 2003), which obscures the simple single loop behavior.
In  weaker flares, on the other hand, this behavior is lost in the noise.
{\it RHESSI} has detected a strong X-class flare (Nov. 3, 2003)
consisting of a single
loop which shows exactly this behavior (Liu \ea 2004, see Fig. \ref{model},
middle). Similar motion but only for the LT source has been seen in
several other {\it RHESSI} flares (see \eg Sui \& Holman 2003).
In rare cases, specially when the FP sources are weak or occulted by the sun,
one can see a double LT source (Fig. \ref{model} right, from Liu 2007,
see also Sui \ea 2005) as envisioned by the cartoon on the left.

\begin{figure}[hbtp]
\leavevmode
\centerline{
\hspace{0.7cm}
\includegraphics[width=5.5cm]{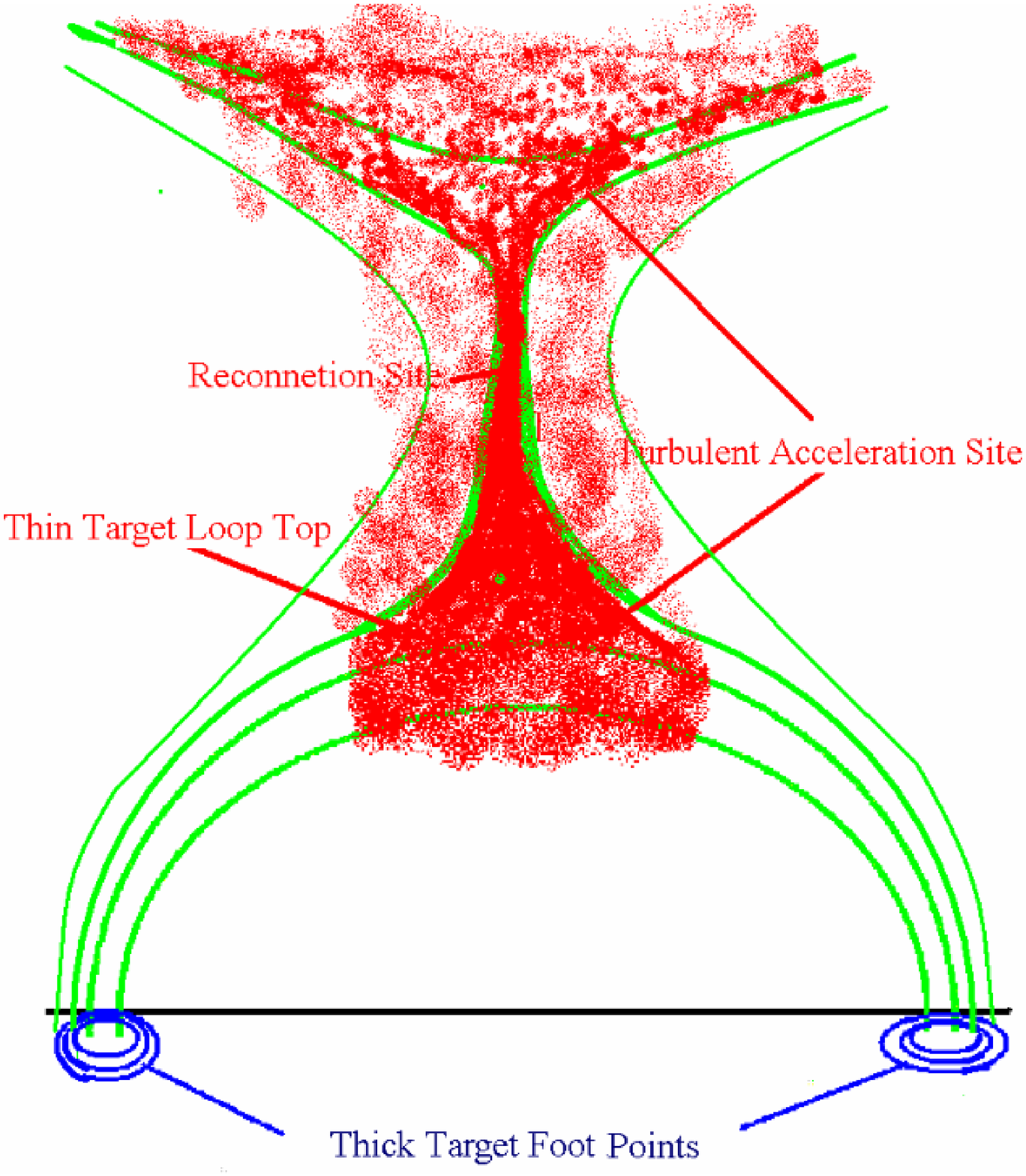}
\hspace{-0.5cm}
\includegraphics[height=6.0cm, origin=c, angle=90]{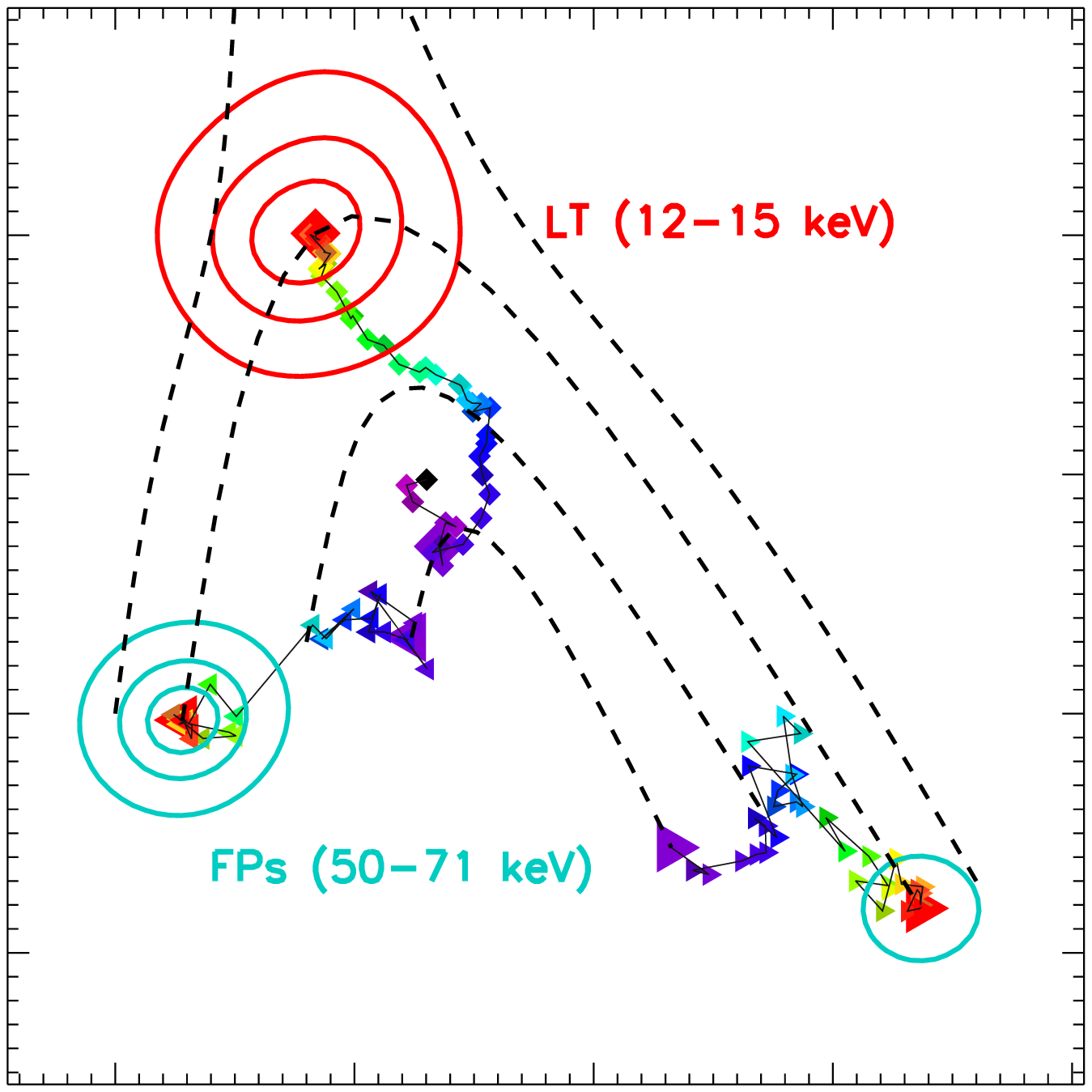}
\hspace{-1.0cm}
\includegraphics[width=6.7cm,height=6.0cm]{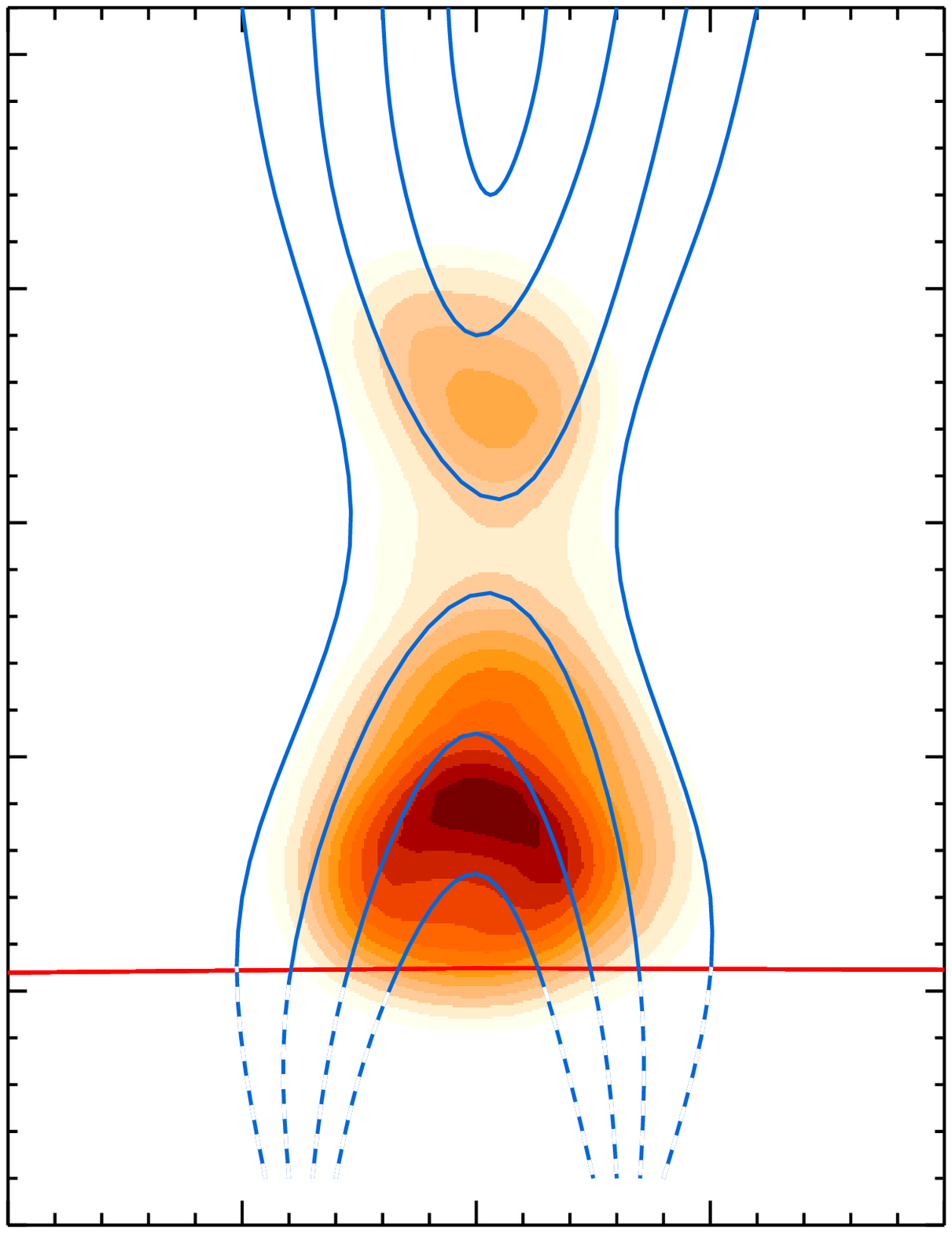}
}
\caption{\scriptsize
{\bf Left panel:} A schematic representation of the reconnecting field forming
closed
loops and open coronal field lines. The red foam represents PWT. {\bf Middle
panel:}
Temporal evolution of LT and FP HXR sources of Nov. 3, 2003 flare.
The symbols indicate the source centroids and the colors show the time with a 20
sec
interval, starting from black (09:46:20 UT) and ending at red (10:01:00 UT) with
contours for the last time. The curves connect schematically
the FPs and the LT sources for three different times showing
the expected evolution for the model at the left; from Liu \ea (2004). {\bf
Right panel:}
Image of flare of April 30, 2002, with occulted FPs
showing an elongated LT source with two distinct peaks as expected from
the model in the left (see Liu et al. 2008). The curves representing the
magnetic lines (added by
hand)
show the presumed occulted FPs below the limb (red line); from Liu (2007).
}
\label{model}
\end{figure}

{\bf Spectral Characteristics:}
Observations over a broad energy range (\eg Marschh\"{a}user et al.  1994,
Dingus et al. 1994, and
Park, Petrosian, \& Schwartz 1997) show several spectral breaks. Petrosian et
al. (1994) have shown that these features
present in the so-called
electron-dominated flares
cannot be due to transport or optical depth effects
 and are natural
consequences of the SA by PWT (Park et al. 1997).
Early high resolution observations (Lin et al.  1981)
showed spectral steepening below a few tens of keV, which is also consistent
with the SA model
(Hamilton \& Petrosian 1992). {\it RHESSI}'s imaging spectroscopy has clarified
this situation
considerably. In general, during the impulsive phase the spectra are dominated
by a non-thermal
component, with the LT always softer than the FPs, and there is a
quasi-thermal component (mostly from the LT)
occurring sometimes before the impulsive phase and always
during the decay phase, which is in concordance with the features of the
accelerated electrons described above.
Figure \ref{ionspectra} shows these observed characteristics of the LT and FP
sources 
for two typical {\it RHESSI} x-ray flares
fitted with theoretical spectra from the SA model.

\begin{figure}[hbtp]
\leavevmode
\centerline{
\includegraphics[width=7.5cm, height=5.5cm]{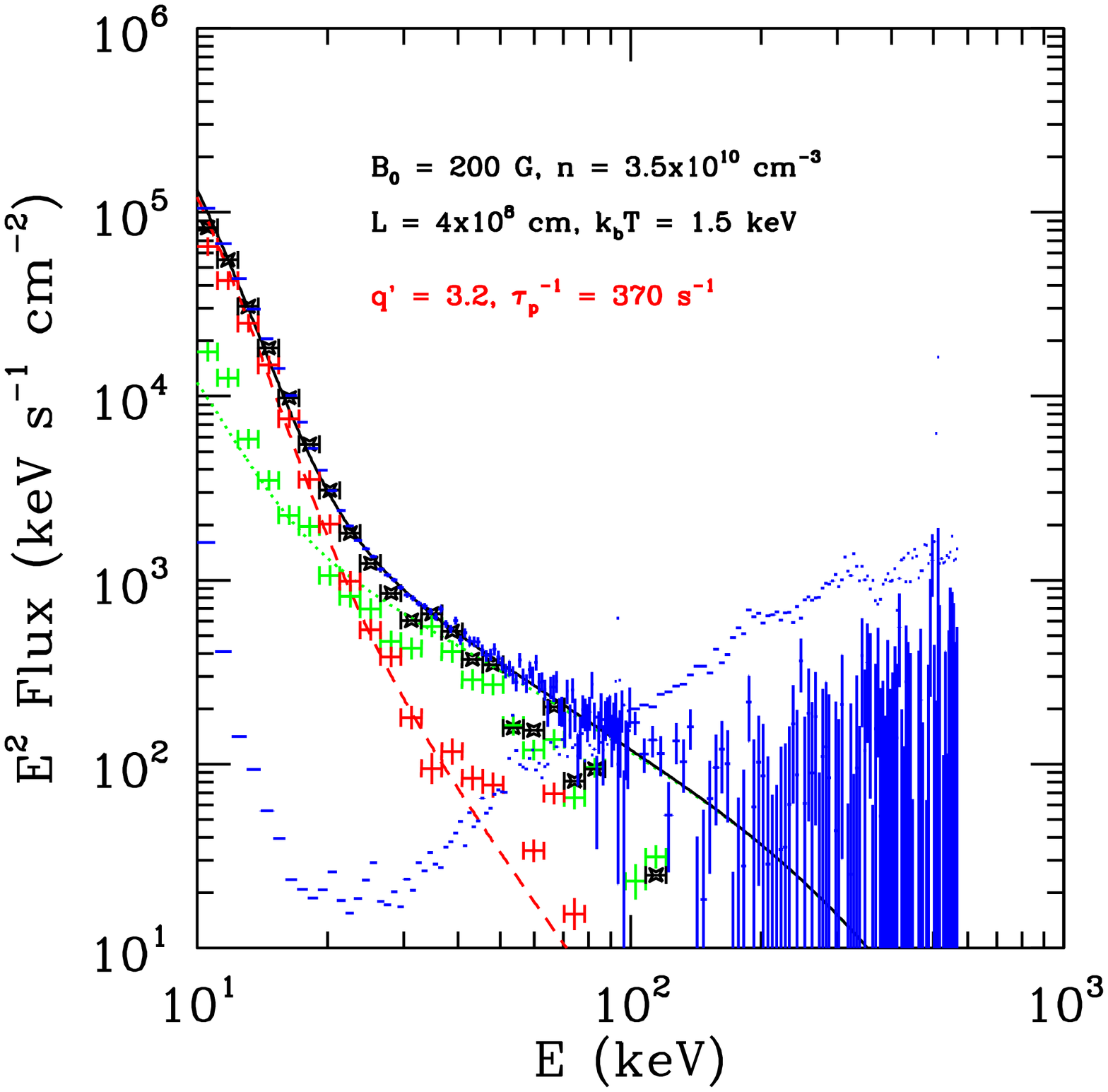}
\includegraphics[width=7.5cm, height=5.5cm]{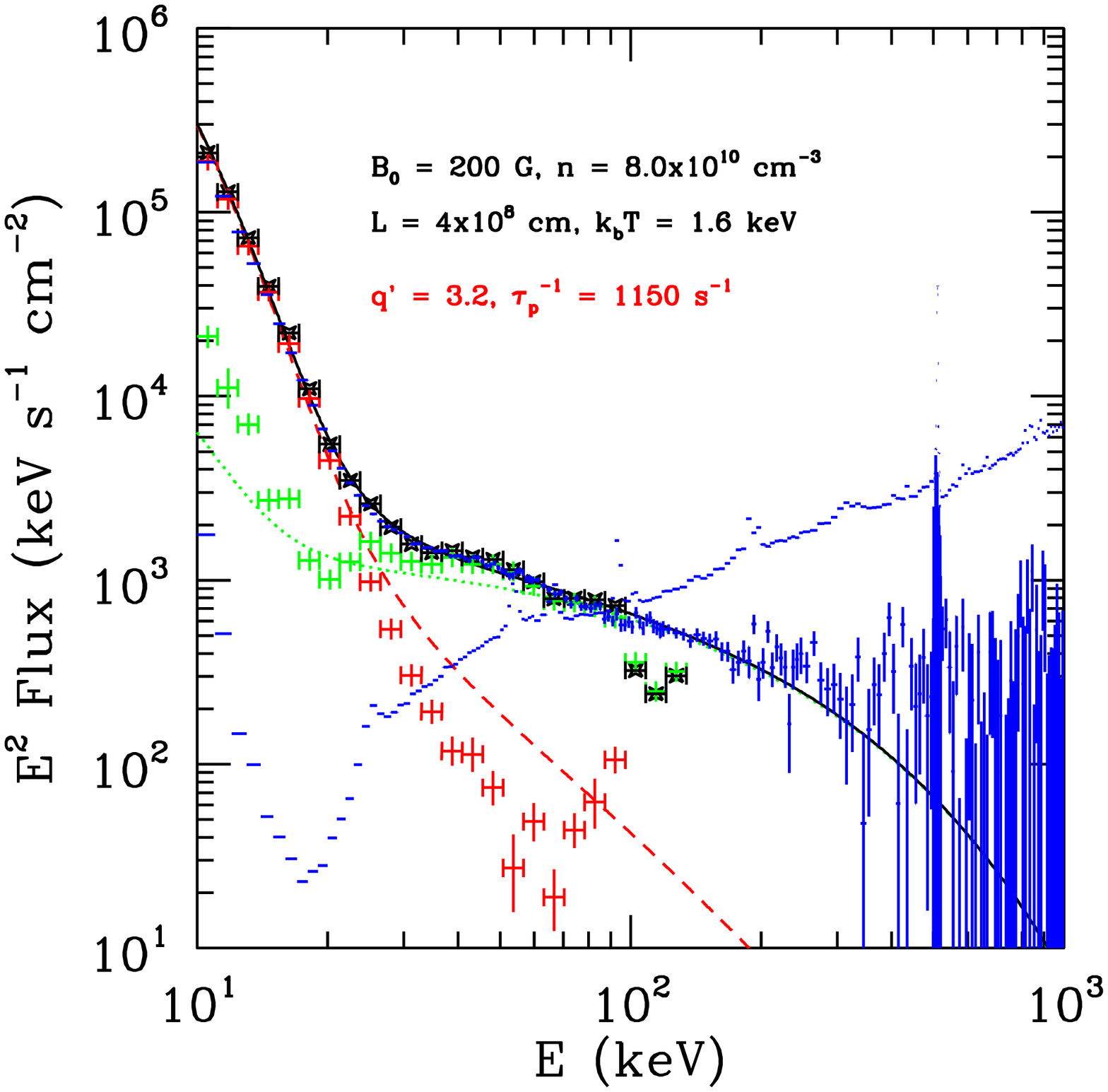}
}
\caption{\scriptsize
{\bf Left panel:} A fit to the total (black), FP (green) and LT (red) spectra of
the 
first HXR pulse of the Sep.
20th 2002 flare observed by {\it RHESSI}. Model parameters are indicated in the
figure. 
The dashed and dotted
lines are model spectra from equations (\ref{kineq}) and (\ref{thick}) showing
the 
presence of a
quasi-thermal (LT) and a non-thermal (mostly FPs) component. The solid line
gives the 
sum of the two. The blue
dashes indicate the level of the background radiation.
{\bf Right panel:} Same as the left panel but for the second HXR peak. Note
that 
compared with the model parameters
for the first HXR pulse, both the gas density and the turbulence energy density 
increase during the second pulse,
resulting more thermal emission and harder photon spectra.
} 
\label{ionspectra}
\end{figure}

{\bf Spectral Evolution:} It has also been known for
sometime that some observations disagree with the model where  \underline{all}
the released energy
goes directly to non-thermal electrons during {\it the impulsive phase}.
Among these are precursor soft X-ray emission referred to as {\it preheating}
and
a slower than expected temperature decline in the decay phase (see, {\it e.g.}
McTiernan et al.  1993)%
\footnote{The expected Neupert (1968) relation also does not seem to
hold (see Dennis \& Zaro 1993; Veronig et al. 2005).}.
In several {\it RHESSI} limb flares Jiang \ea (2006) find that during the decay
phase the
(resolved) LT
source continues to be confined; it does not extend to the FPs as one would
expect
if the evaporation
fills the whole loop with hot plasmas.
Moreover,  the observed energy decay rate is calculated to be
higher than the radiative cooling rate
but much lower than the (Spitzer 1961) conduction
rate. The latter and the confinement of the LT source require suppression of the
conduction and a continuous energy input.
PWT can be the agent for both. It can reduce the scattering mean
free path and energize the plasma continuously during the decay phase (albeit at
a
diminishing rate).

The observed spectral  breaks and the temporal evolution also agree with the
model results discussed above. During the build-up and decay (\ie pre- and
post-impulsive)
phases when the level of PWT is low it only manages to ``heat" the plasma with
little
acceleration into a non-thermal tail. Particle acceleration begins in earnest
during
the impulsive phase when the rate  of reconnection and production of PWT rises
and then falls
after reaching high levels. During this rise and fall the spectra undergo
soft-hard-soft
evolution as observed in most flares.

\begin{figure}[hbtp]
\leavevmode
\centerline{
\includegraphics[width=7.0cm]{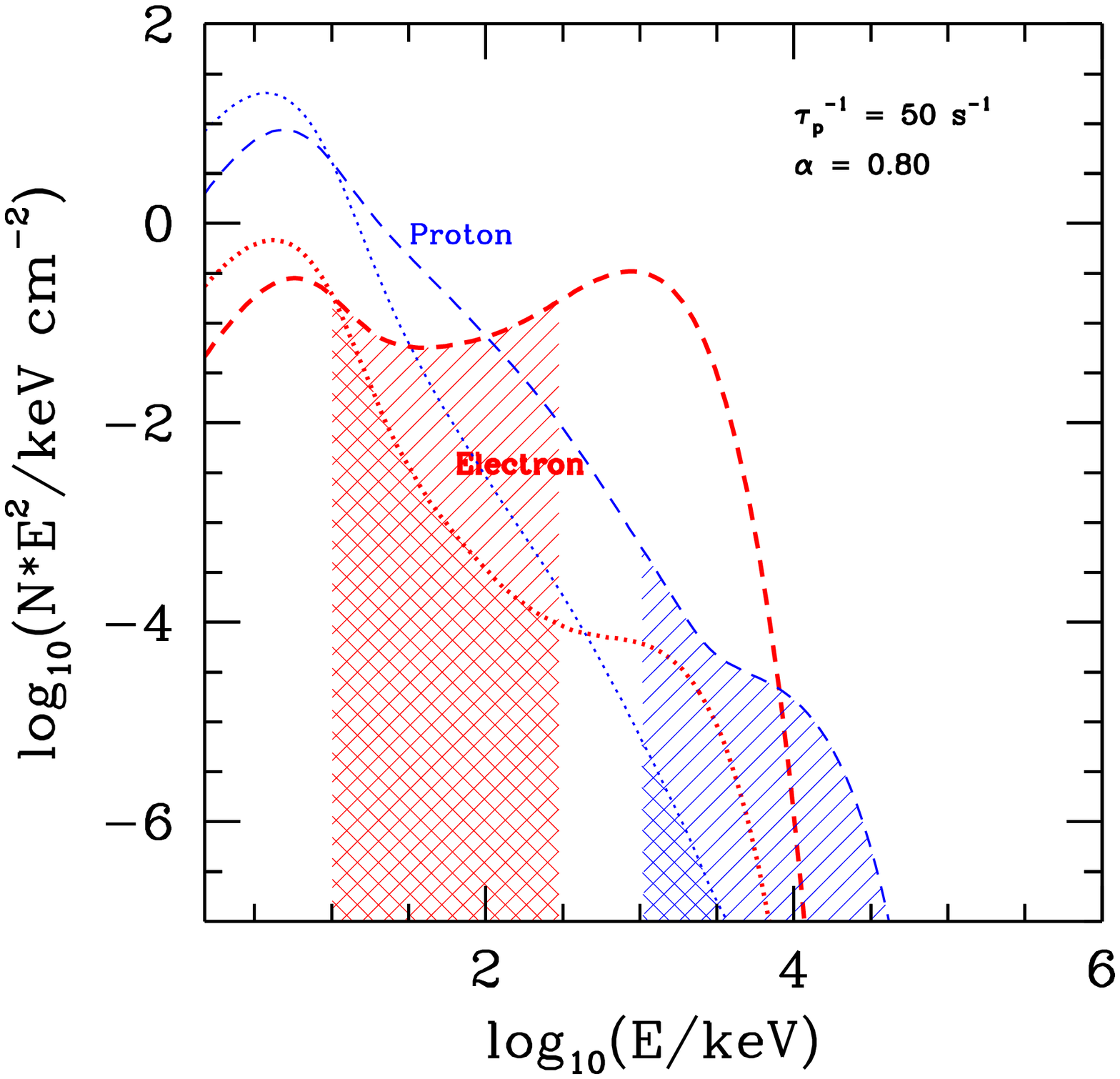}
\includegraphics[width=7.0cm]{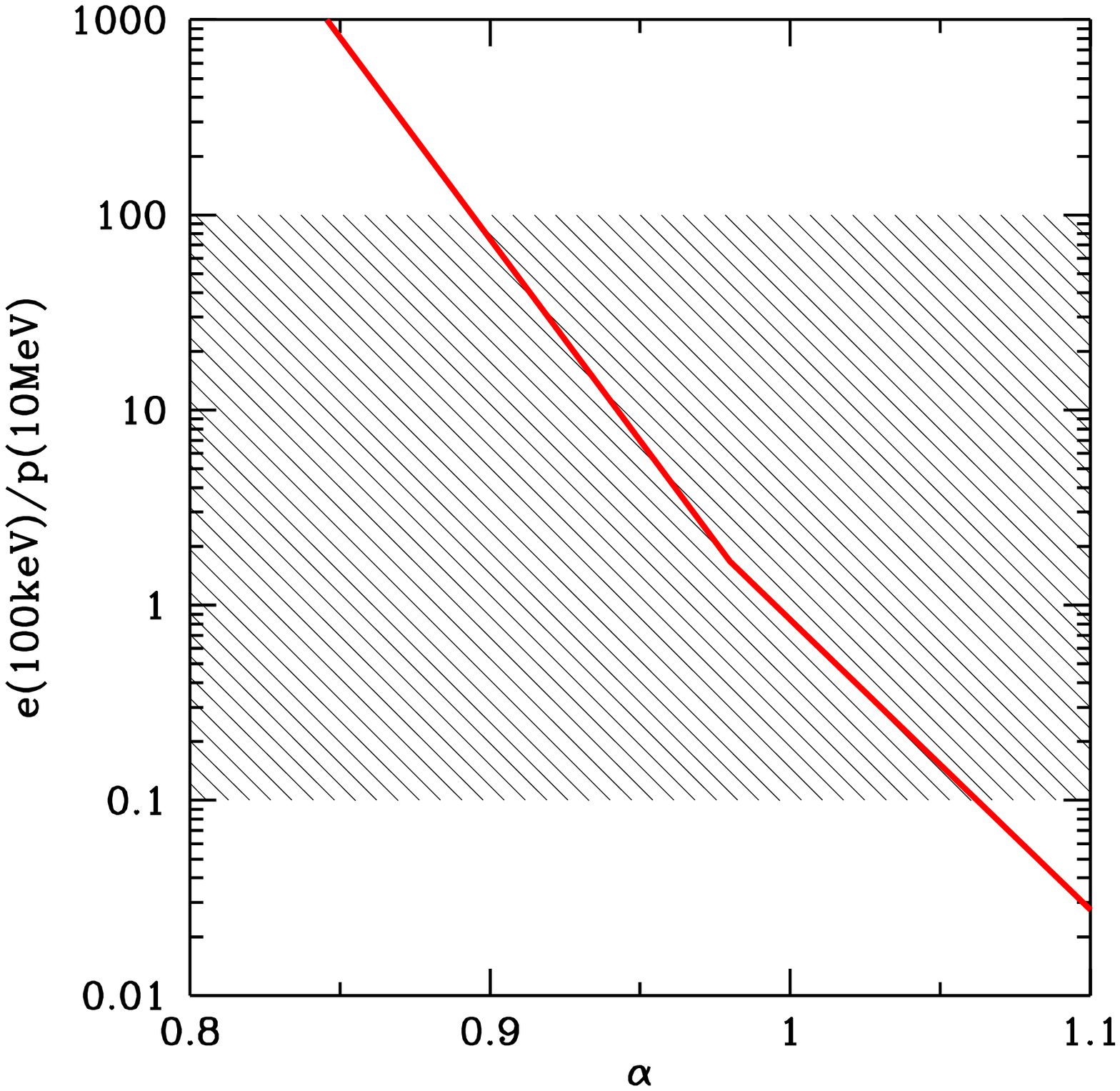}
}
\caption{\scriptsize
{\bf Left panel:}
Same as left panel of Figure \ref{spectra} but for a different $\alpha$ showing
the dependence of relative electro and proton spectra on $\alpha$.
{\bf Right panel:} Variation with $\alpha$ of the ratio of the energies of the
accelerated electrons and protons in the spectral energy range shown by the
shade region in the left panel.
}
\label{epratio}
\vspace {-0.5cm}
\end{figure}

\subsubsection{Radiative Signatures of Protons and Ions}

Observations  of gamma-ray lines in larger solar flares provide
evidence for acceleration of protons and ions. The SA model has been
the working hypothesis
here as well  (see works by Ramaty and collaborators earlier, and Murphy and
Share recently). For these
flares
the ratio of energy content of accelerated electrons to
protons
(in the hatched range of Fig. \ref{spectra}, left)
ranges from 100 to 0.03
(Mandzhavidze \& Ramaty 1996);  this ratio is  larger in electron dominated
flares.
On the other hand, simple theoretical models, for example
acceleration by Alfv\'{e}n waves commonly used, favor a much smaller ratio.
PL04  show that with a more complete treatment of the
wave-particle interaction a barrier appears for proton acceleration,
which reduces the number of accelerated protons dramatically,
in a way that the above ratio is very sensitive to the plasma parameter
$\alpha$ in equation (\ref{alpha}).
Figure \ref{epratio} shows this dependence. The left panel is same as the left
panel in Figure \ref{spectra} but for a smaller $\alpha$. The right panel shows
the variation with $\alpha$ of the ratio of the energy of electron to protons
(in the respective shade region).
One consequence of this is
that the proton acceleration will be more efficient in larger loops, where the
$B$ field is
weaker, and at late phases, when  due to evaporation $n$ is higher.
This can explain the offset of the centroid of the gamma-ray line emissions from
that of HXRs seen by {\it RHESSI} (Hurford et al.  2003).
It can also account for the observed delay of nuclear line emissions relative
to HXR emissions seen in some flares (Chupp et al.  1990).
This trend is also true for other ions because the acceleration barrier
moves to lower energies with decreasing $\alpha$.
or
In general, higher densities, lower magnetic fields favor
acceleration of ions
versus the electrons and can explain the diversity in their relative observed
values both at the flare site and in the interplanetary space.

\subsubsection{SEP Spectra and Abundances}

It is commonly believed that the observed relative abundances of ions in SEPs
favor a SA model
({\it e.g.}
Mason et al. 1986 and Mazur et al. 1992). More recent observations have
confirmed
this picture (see
Mason et al. 2000, 2002, Reames et al. 1994 and 1997, and Miller 2003).
There are similarities and
differences among the spectra and abundances of different ions
from event to event.
One of the most vexing problem of SEPs has been the  enhancement of $^3$He
in the so-called {\it impulsive events}, which sometimes can be $3-4$ orders of
magnitude above the
photospheric value. As stated at the outset these events, which we prefer to
call
\3he-rich rather than impulsive, is the focus of thees proceedings.  There has 
been many attempts to  explain this
and the related enhancements of heavy ions observations. This will be the
subject of the next sections.

\section{Enhancement of \3he and Heavy Ions}

As stressed in the previous chapters one of the most puzzling aspects of
observations of SEPs is the enhancement of \3he and heavy ions  relative to
their photospheric
values in many of the so-called  impulsive events. Over the years there has been
many attempts to account for these observations. In this section we first
present a
brief
review of past attempts at tackling this problem and then describe how well the
SA
model described above can account for these aspects of solar flares.

\subsection{Past Works}

Most of the proposed models, except the  Ramaty and Kozlovsky (1974) model based
on  spalation (which has many problems), rely on resonant wave-particle
interactions
to produce the observed enhancement. The  earliest theoretical
models addressing this issue are those of Ibragimov and Kocharov (1977) and Fisk
(1978).

\begin{itemize}

\item

Ibragimov and Kocharov claim that under the influence of Langmuir turbulence the
plasma ion temperature will increase linearly in time with a rate proportional
to the momentum diffusion coefficient. Citing another paper they also claim that
the diffusion rate will be proportional to the Coulomb charge times the
gyrofrequency which favors heating of \3he relative to protons and \he4. A
second stage of acceleration with a velocity or energy threshold for all
particles will then accelerate a larger fraction  of \3he ions than \he4 and
proton. The parameters of the model can be adjusted to obtain the observed
enhancements.
\item

Fisk's model relies on the fact that  \3he  can interact more efficiently
than
\he4 with (electrostatic hydrogen-cyclotron) waves having frequency between
proton and \he4 gyrofrequencies. Damping of these waves preferentially heats
\3he  ions. There are two major shortcomings in this model. The main
difficulty
here is that to have sufficient overheating one requires excitation of waves
close to \3he  gyrofrequency (=4/3 of \he4) which can be produced if electron
to
proton temperature ratio and/or the \he4 to hydrogen ratio are large. Again,
this model does not answer the full question of production of enhanced
population of non-thermal \3he ions. It leaves this to a secondary process which
also
must satisfy the requirement of having a threshold velocity for acceleration
near
the thermal velocity of the heated \3he ions.  Fisk also
tries to explain the enhancement of the heavy elements the same way but relying
on the resonant interactions with the second gyro-harmonics. This aspect also
may not be correct.

\item

Temerin and Roth (1992) use essentially same idea but with
electromagnetic hydrogen-cyclotron waves. They carry out numerical simulation
based on somewhat artificial model (waves moving down the variable field lines
and \3he  ions moving up that interact when the resonant condition is
satisfied).
 They also produce some heating of \3he  ions but do not discuss the spectrum
nor
the expected ratio of \3he/\he4. Again, like  in the Fisk model first  and
second harmonics
($n=1$ and 2) are invoked for \3he  and heavy ion heating, respectively.
However, there is no estimate of heavy ion fluxes. Curiously, Fisk is not
acknowledged except in passing mentioning only the shortcoming  mentioned above.

\item

Miller and Vi\~nas (1993) expand on Temerin and Roth work by
inclusion of more waves and more-realistic plasma conditions. Most of this
paper is devoted to calculating the growth rate  and dispersion relation of
waves excited by a thermal, isotropic ``beam" of electrons superimposed as  a
bump in a somewhat cooler thermal plasma consisting of electrons,
protons and alpha particles. The acceleration of particles is carried out using
orbit calculations for test particles. As in the above works it merely leads to
some heating of iron and \3he  ions. It is then stated that this heating is
more
than what they expect to occur for other ions (\eg CNO) and \he4, respectively.
The required ``beam" densities are much higher than those required for
describing
the Type III radio emission. (These authors provide a good summary and analysis
of earlier works).

\item

Zhang (1995) uses a more refined version of  the Fisk model and evaluates the
degree of overheating of \3he ions relative to \he4. In addition he includes a
simple calculation of subsequent acceleration by a ``Fermi" process and derives
a general spectrum. However, these spectra are not compared with the observed
ones; they are essentially exponential in form and will not fit well to the
observations. Instead he evaluates the abundance ratios and plots these versus
some important parameters which can be compared with the gross features of the
observations.

\item

Paesold, Kallenbach and Benz (2003)  rely on a somewhat different
aspect of the acceleration process. In their model
waves
are excited by anisotropic distribution of electrons (so-called Fire Hose
Instability) with a growth rate  that is slightly higher at the \3he
gyrofrequency
than that of \he4. It is then claimed that this can lead to a much larger
(exponential)
difference in the wave energy density at the corresponding frequencies. With a
larger wave density at its disposal \3he  acceleration proceed more rapidly. The
acceleration method is used is the same as in Miller and Vi\~nas discussed
above. However, unlike the above works this paper addresses the spectral
differences between \3he  and \he4, but there is no comparison with actual
observed spectra. This model also predicts that there should be i) many more
accelerated protons (which is swept under the rug invoking some energy
limitation) and ii) a \underline{suppressions} of iron and other heavier ions
(whose
 $Q/A$  is less than that of \he4 or CNO) instead of the observed
enhancements.

\end{itemize}

In summary the common theme among all these models is that they use the
difference between the $Q/A$ of \3he and \he4 and require some non-Maxwellian
electron distribution to excite the waves that then preferentially heat
\3he ions. Some also require very special conditions.
The outcomes of all these works are in general some differential heating with
the actual acceleration
often relegated to a subsequent process. There is little or no comparison with
observed spectra and every model requires some additional factor to explain the
enhancements observed for iron and heavier ions.

In contrast to this the model proposed by Liu, Petrosian and Mason (2004, 2006)
based on the SA by PWT (PL04) treats heating and acceleration as a single
process, can
reproduce some of the observed spectral features, and can account for the range
of the observed
enhancement. On the other hand, at this stage of its development this model does
not address the details of the production of the waves or turbulence and relies
on injection, presumably during the reconnection process,  and the subsequent
cascade and damping of the waves. Some of the salient results of this model are
described next.

\subsection{Enhancements and Stochastic Acceleration}

In this section we describe the role the SA model described in the previous
sections can play in the enrichment of \3he and other elements. The formalism
described above can be applied to the acceleration of all ions in the same way
used to fit the radiative signatures of the accelerated electrons and protons to
observations. All the results presented below are based on acceleration by PPWs
and most of the results are from Liu \ea (2004, 2006).

\begin{figure}[hbtp]
\centerline{
\includegraphics[height=7.5cm,angle=0]{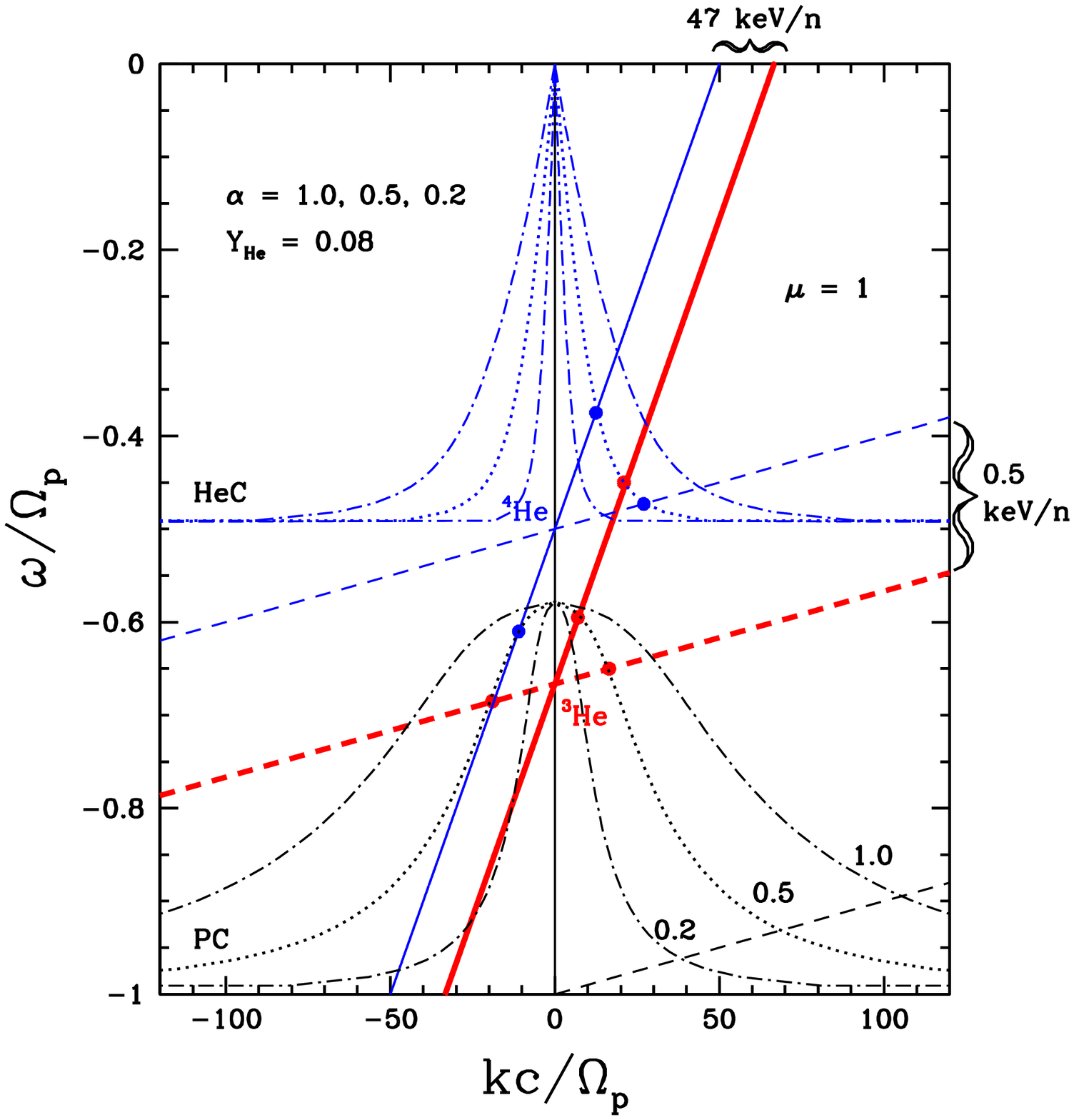}
\includegraphics[height=7.5cm,angle=0]{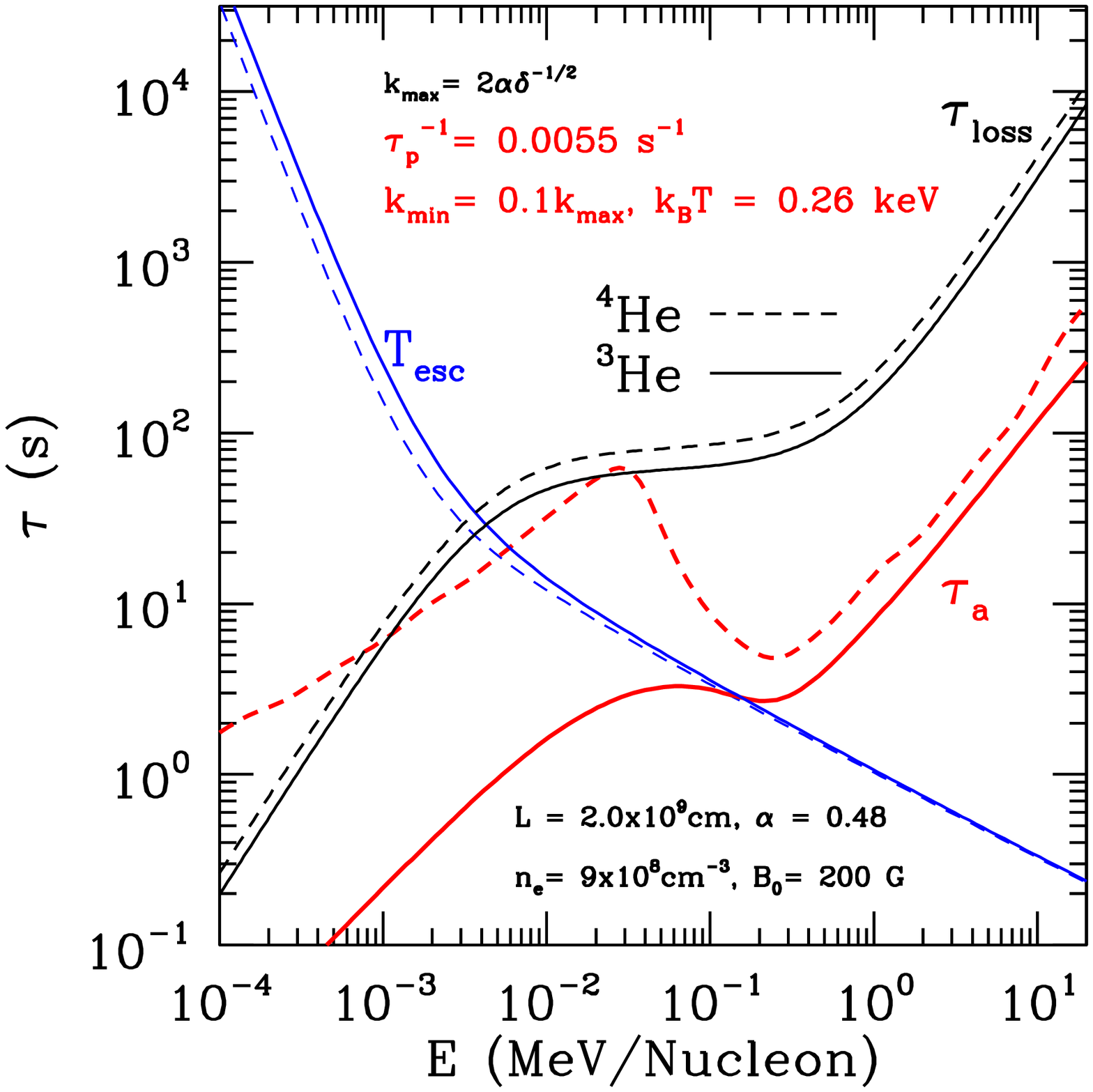}
}
\caption
{\scriptsize
{\bf Left panel:} Dispersion relations for parallel propagating waves and
resonance conditions for \3he and \he4 for the specified energies and three
different values of $\alpha$.  Note that the resonance lines for \3he start (at
$k$=0) at its gyrofrequency (divided by $\gamma\sim 1$), which is larger (in
absolute value) than that of \he4. As a result at low energies \3he  has more
resonant intersections (filled circles)  than \he4. {Right panel:} Acceleration,
escape and energy time scales
for \3he an \he4. The main difference here is the shorter acceleration time for
\3he at lower energies because of its stronger interactions with waves.
}
\label{Hetimes}
\vspace{0.5cm}
\end{figure}

Figure \ref{Hetimes} (left) shows the dispersion relation near the proton and
\he4 gyrofrequencies and the resonant conditions for two different energies of
\3he and \he4 ions. The presence of fully ionized \he4 (about 8\% of protons) in
the background plasma divides the Alfv\'en wave-proton cyclotron branch into two
distinct modes which we have labeled proton cyclotron (PC) and He cyclotron
(HeC) branches. Because of the gap between these two modes at low values of
the wavevector $k$,  low energy \he4 ions resonate only with the HeC branch
while
\3he ions can resonate with both branches and are
accelerated more efficiently. The right panel of this figure depicts the
relevant
time scales showing the much shorter acceleration time for \3he than \he4 at low
energies. This is the primary cause of the \3he enrichment.

As a result of this more efficient acceleration most \3he ions are accelerated
to form a non-thermal distribution. While \he4 ions are heated up forming a low
energy quasi-thermal bump with only a small fraction reaching to high energies.
Figure \ref{Hespectra} shows model spectra fitted to two \3he rich SEP events
displaying the above features.  Another result of
the efficient acceleration is that one expects a high degree of acceleration of
\3he even in smaller events such that the \3he/\he4 abundance ratio is expected
to decrease with increasing fluence of the event. This is what is seen (see Fig.
XXX in Chapter 2) and is demonstrated in the right panel of Figure
\ref{Hespectra} where we show spectra for three different values of the
acceleration rate $\tau_p^{-1}$. For higher rates more of the \he4 ions in the
quasi-thermal tail reach into the non-thermal tail decreasing the \3he
enrichment and the \3he/\he4 ratio.

\begin{figure}[hbtp]
\centerline{
\includegraphics[width=5.8cm,angle=0]{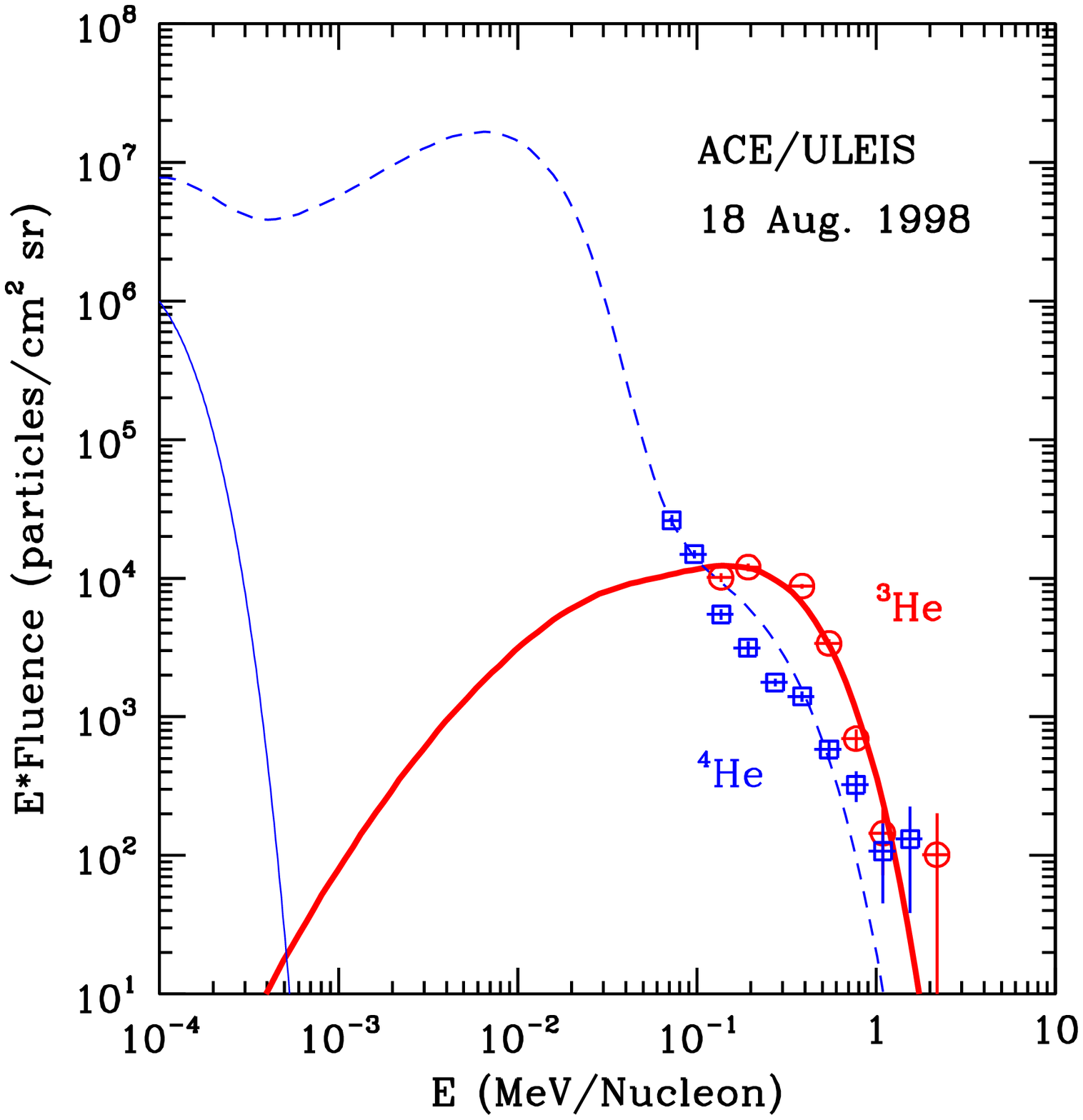}
\hspace{-0.5cm}
\includegraphics[width=5.8cm,angle=0]{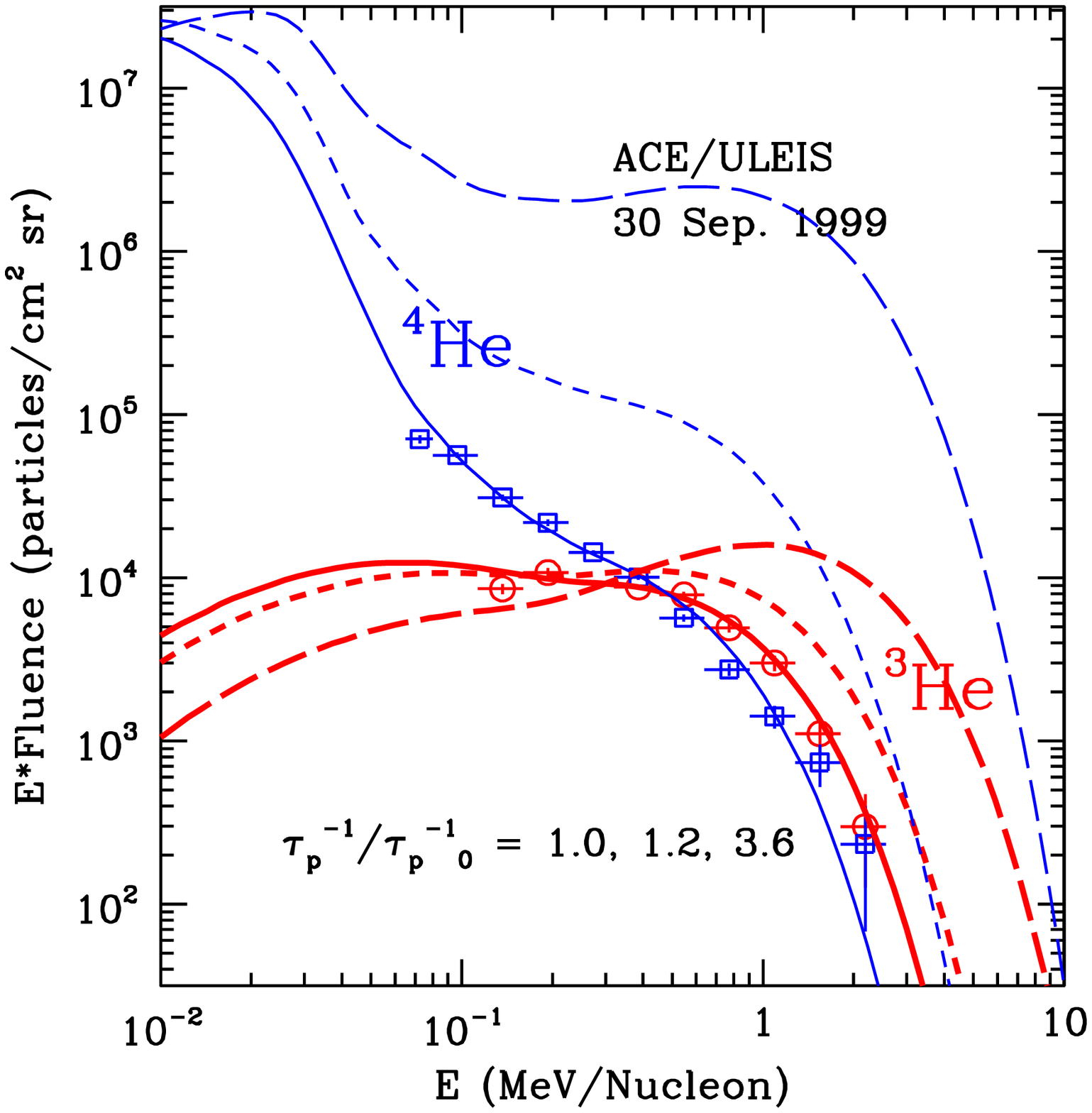}
\hspace{-0.5cm}
\includegraphics[width=5.8cm,angle=0]{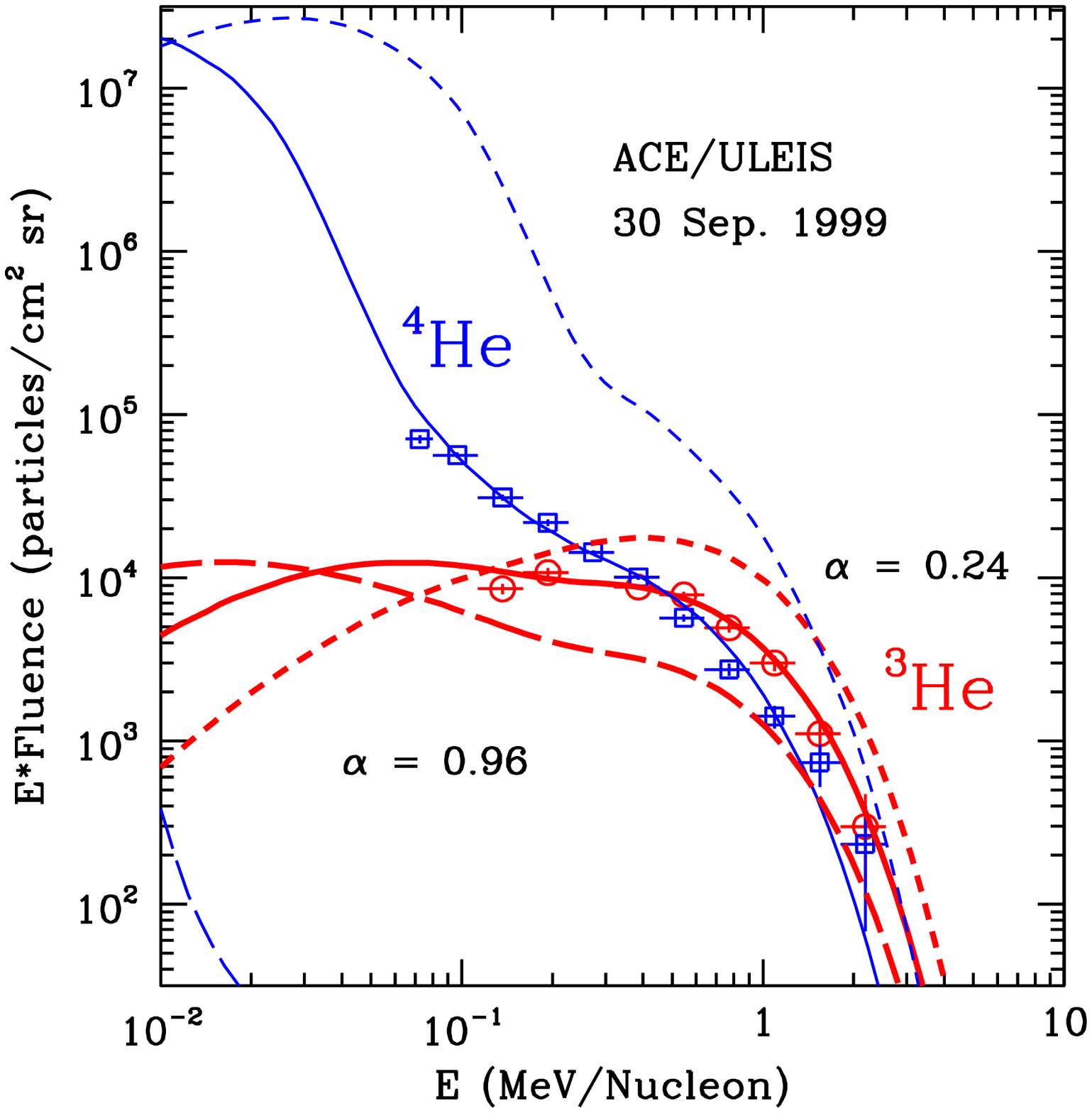}
}
\caption
{\scriptsize Fits to \3he and \he4 spectra of two \3he rich SEP events. Note
that because of the shorter acceleration time of \3he at low energies most of
the \3he is accelerated into super thermal tail while most of \he4 ions are
heated up to a high 'temperature' quasi-thermal distribution with only a smaller
fraction
forming a non-thermal tail. The middle and right panels also show two additional
sets of
model spectra with somewhat higher and lower values of $\tau_p^{-1}$, which is
proportional to the turbulence energy density or the rate of acceleration,  and
$\alpha$, respectively.
}
\label{Hespectra}
\vspace{0.5cm}
\end{figure}

The dependence of \3he enrichment on model parameters is further 
demonstrated in Figure \ref{Hetaup}. The most influential parameter is the
acceleration rate $\tau_p^{-1}\propto {\cal W}$ and there is some variation with
$\alpha$ at low
acceleration rates. There is little dependence on the (unknown) spectral
parameters of the turbulence. The trend and the range of the
variations presented here are in agreement with the observations described in
chapter 2. It remains to be seen whether or not the spectral trend presented in
Figure \ref{Hespectra} (middle and right) also agrees with observations. It
should also be possible
to use the observed distributions of \he4 and \3he fluences or peak fluxes to
further test this promising model and constrain its parameters. This will
require a sample of events with known selection criteria and observational
biases.

\begin{figure}[hbtp]
\centerline{
\includegraphics[height=7.0cm,angle=0]{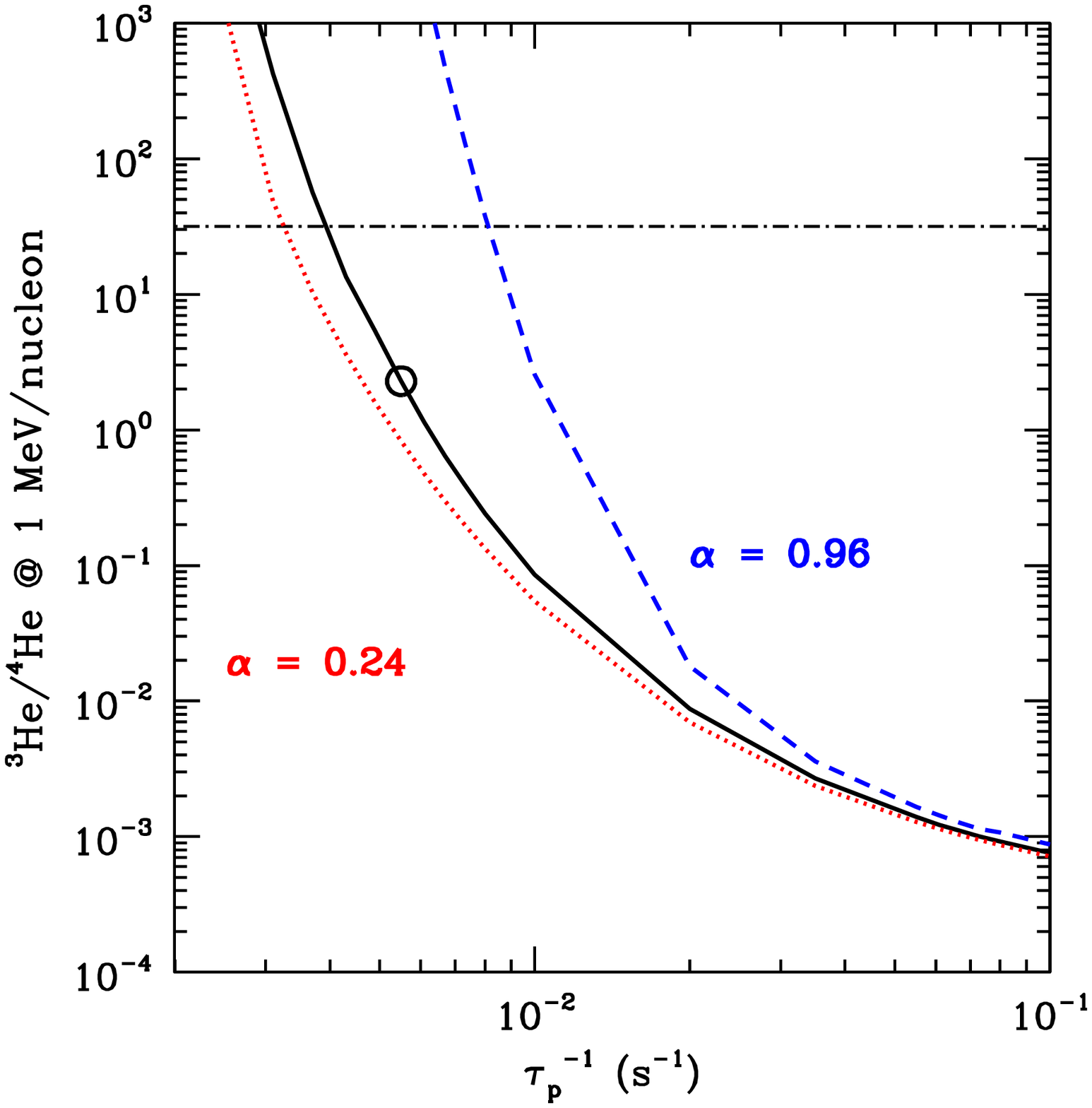}
\includegraphics[height=7.0cm,angle=0]{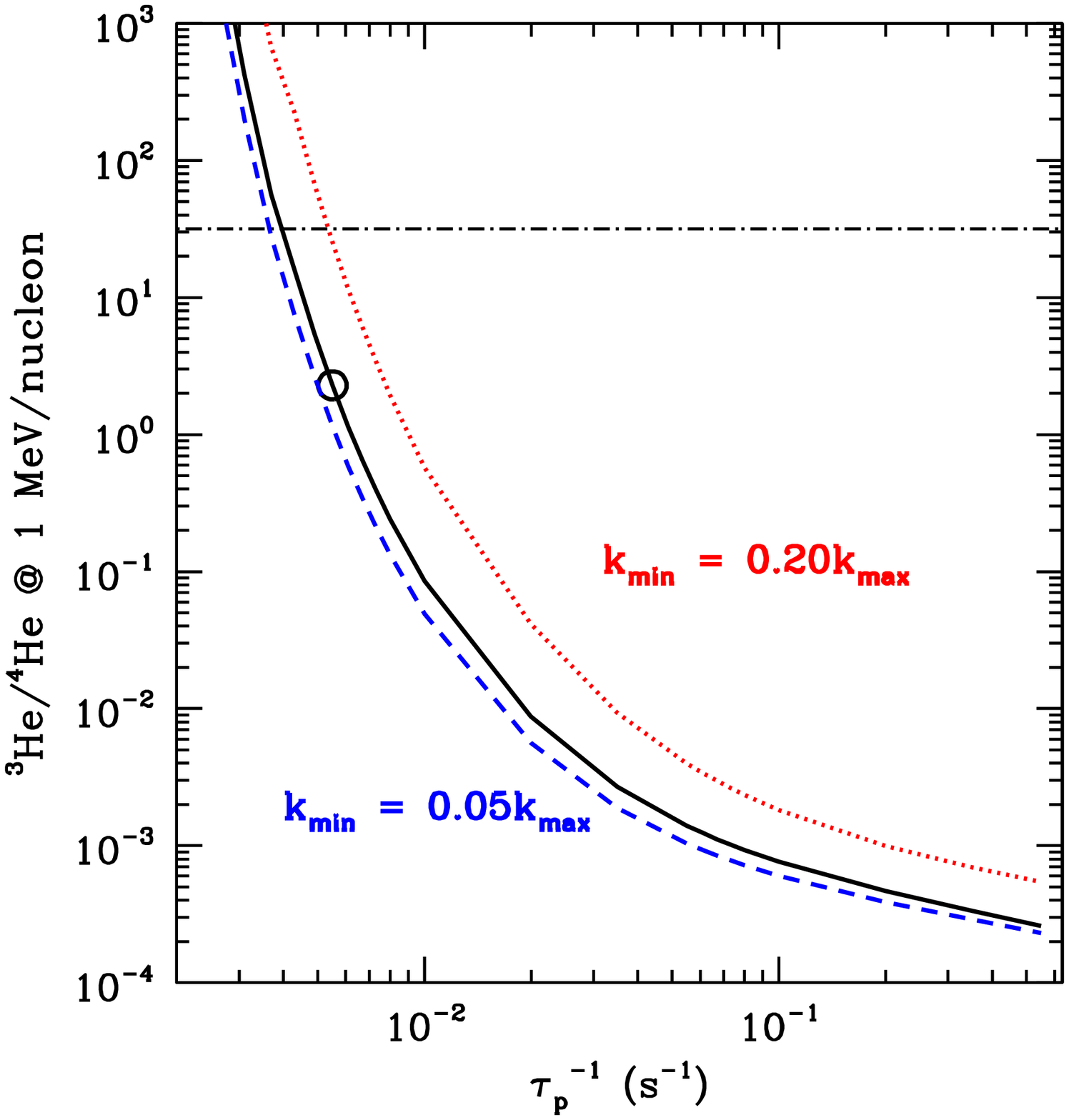}
}
\caption
{\scriptsize Variation with acceleration rate (or level of turbulence) of the
\3he/\he4 fluence ratio at 2 MeV per nucleon for different values of $\alpha$
(left) and the
spectral break wave vector $k_\max$ (right). The black lines are for best fit
values of $\alpha=0.46$ and  $k_\min=0.1 k_\max$ for the 30 Sep. 1999 event
shown
in Figure \ref{Hespectra} (middle and left). These curves  show the possibility
of
obtaining a large range of the ratio of \3he/\he4 and that the main
characteristics of turbulence affecting this ratio is the energy density ${\cal
W}\propto \tau_p^{-1}$ with smaller effect due to $\alpha$ and a relatively
insignificant effect due to unknown details of the turbulence spectrum (see Liu
\ea 2006).
}
\label{Hetaup}
\end{figure}

\section{Summary and Discussion}

We have discussed possible mechanism of  accelerations in solar flares and SEPs,
and pointed out that in all
mechanisms plasma  waves and turbulence play a major role. We have then argued
that in solar flares the most likely mechanism is stochastic acceleration  by 
plasma  waves and turbulence and have demonstrated
that the predictions of this model are consistent with most of the observed
flare
radiative signatures. We have also shown that the observed \3he enrichment and
the spectra of \3he and \he4 can  be described by such a model without
requiring any special conditions. The same model that successfully describes the
observed radiation can also account for this long standing puzzle. Unlike
previous models, which addressed only the enhancement (without detailed
comparison with observed spectra) and required  special conditions and/or two
stage acceleration, the  model presented here neither requires special
conditions nor another acceleration mechanism.

Another important problem of SEP observations is enhancement of abundances of
heavier ions with the enhancement increasing with ionic mass or mass-to-charge
ratio described in Chapter 2. Here the situation is more complicated because of
the
uncertainty about the charge state of these ions and the observations of
dependence of the charge state on the ion energy discussed in previous chapters.
These observations also favor
acceleration in relatively high densities of the lower corona, where the
stochastic acceleration 
invoked here occurs. Figure  \ref{heavies} (left)  shows abundances of a large
number of ion species obtained from a
composite of many SEP events (red points, Mason, private communication). The
black points and the line show
our preliminary attempt (Liu \& Petrosian in preparation) to fit these
observations. Fits to protons \3he and
\he4 are as above and in plotting the line we have assumed a charge of 16 for
all the elements above iron.

\begin{figure}[hbtp]
\centerline{
\includegraphics[height=7.0cm,angle=0]{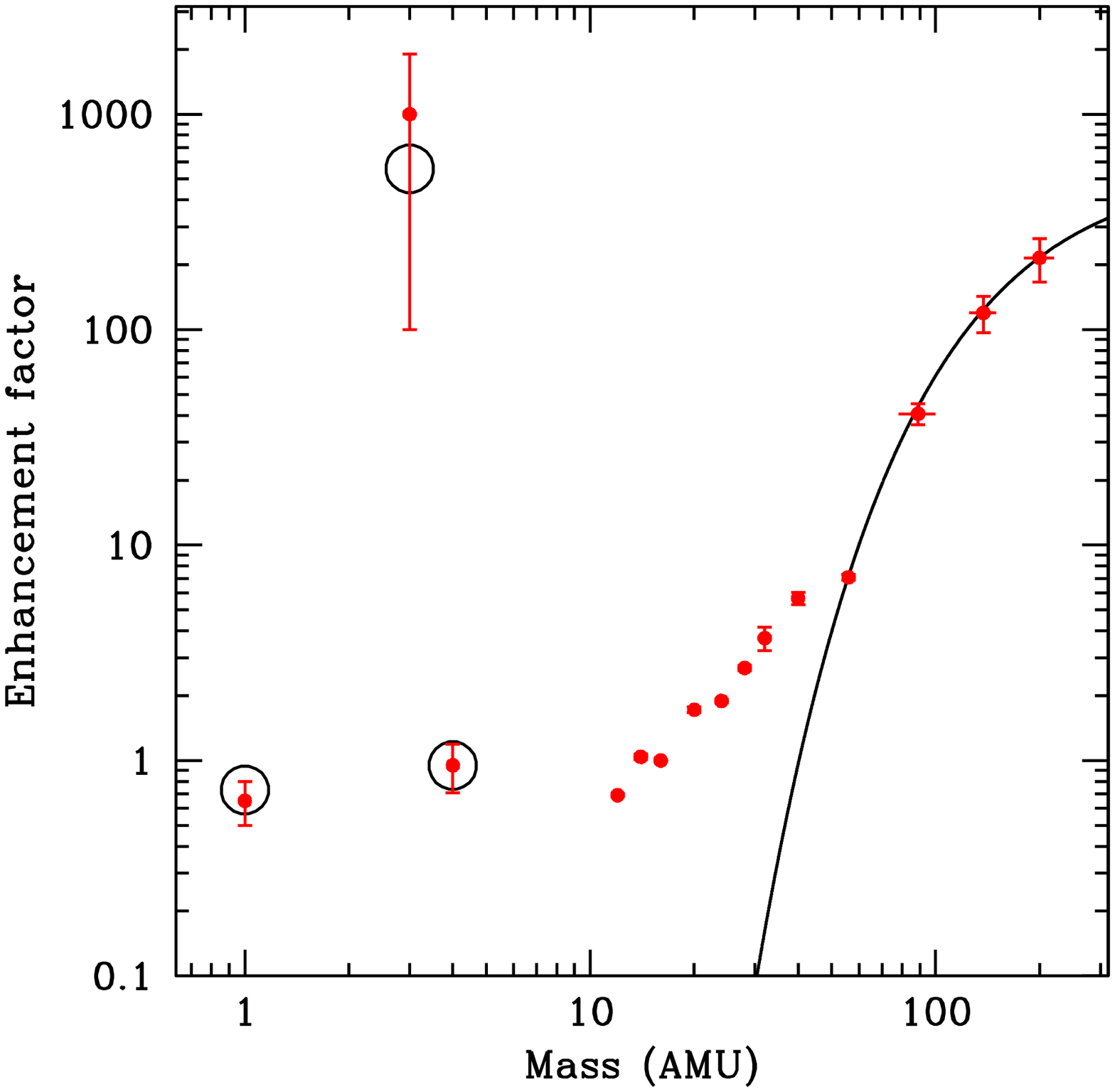}
\includegraphics[height=7.0cm,angle=0]{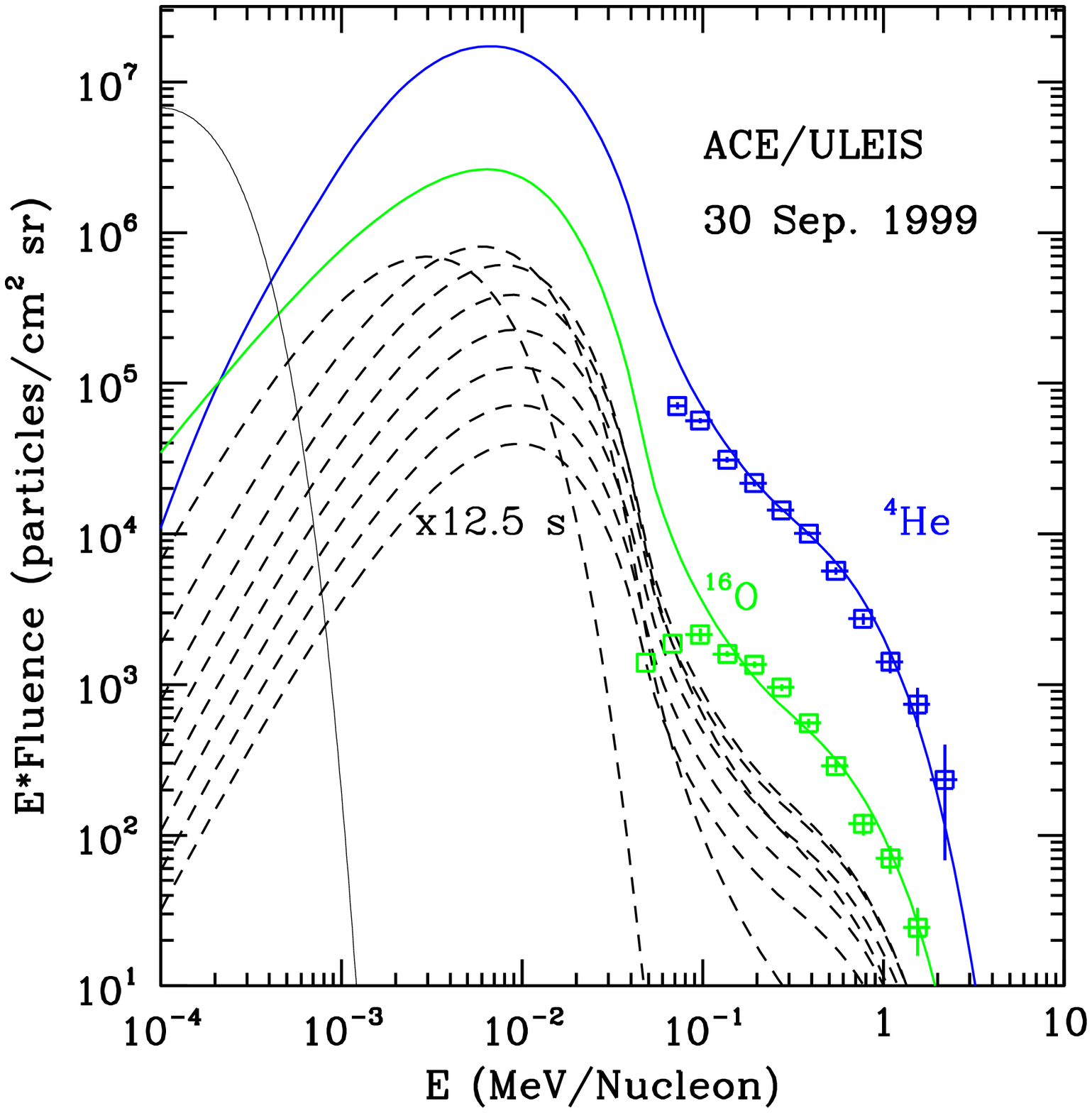}
}
\caption
{\scriptsize {\bf Left panel:} A preliminary fit by the SA model to different
ion enhancements obtained from a composite of many events (red points from
Mason, private communication). The open circles and the black curve are the
model
results assuming a charge of 16 for all elements heavier than iron (Liu \&
Petrosian, in preparation). This, of course, is inappropriate for intermediate
ions where the time-independent model described here cannot describe the
observations. {\bf Right panel:} A preliminary result from a time-dependent
treatment showing the evolution of $^{16}$O (dashed lines) spectrum and the time
integrated spectrum (solid green) fitted to  observed spectrum of fluences of
$^{16}$O and \he4 for a typical event.
}
\label{heavies}
\end{figure}

This scenario in the form presented here will encounter difficulty with the
observations of the intermediate ions (\eg CNO). We note that most of the
observed spectra refer to the event fluence. The models described above are
based on the \underline{time-independent} (or the steady state) equations. This
is valid when the acceleration time is shorter than the dynamic time scale of
the
event (\eg event duration), in which case the steady state results are good
approximation. A more accurate analysis requires a \underline{time-dependent}
analysis. Figure \ref{heavies} (right) shows a preliminary result from such a
model (East, Petrosian \& Liu, in preparation) from such a model. The dashed
lines
show the time evolution of the spectrum of $^{16}$O and the solid green line the
time integrated spectrum appropriate for the fluence spectrum shown by the green
squares. This model seems to provide a reasonable fit (except at low energies
where the uncertainties are large) for both  $^{16}$O  and \he4. 
Clearly more data and more detailed model are
required to put this comparison on the same footing as those described above for
electrons, protons and He ions.

\section{REFERENCES}

\def\et{{\it et~al.}}
\def\aa{{\it Astr. \& Astrophys.}}
\def\anyas{{\it Ann. N.Y. Acad. Sci.}}
\def\apj{{\it Ap.~J.},\,\,}
\def\apjl{{\it Ap.~J. (Letters),\,\,}}
\def\ass{{\it Ap. Space Sci.}}
\def\jgr{{\it J.~Geophys. Res.}}
\def\mnras{{\it M.N.R.A.S.}}
\def\nature{{\it Nature}}
\def\prd{{\it Phys. Rev.~D}}
\def\pra{{\it Phys. Rev.~A}}
\def\ssr{{\it Space Sci. Rev.}}
\def\rmp{{\it Rev. Mod. Phys.}}
\def\sphys{{\it Solar Phys.}}
\def\JL{Leach,~J.\ }
\def\VP{Petrosian,~V.\ }

\def\refer{ \par \noindent \hangindent=2pc \hangafter=1}
\baselineskip = 10 true pt

\refer Amato, E., \& Blasi, P. 2005, MNRAS, 364, 76
\refer Andr\'e, Mats 1985, {\it J. Plasma Phys.,} 33, 1
\refer Boris, J. P., Dawson, J. M., Orens, J. H., \& Roberts, K. V. 1970, 
{\it Phys. Rev. Lett.}, 25, 706
\refer Cassak, P. A., Drake, J. F. \& Shay, M. A. 2006, \apj 644, L145
\refer Chandran, B. 2005, {\it Phys. Rev. Lett.}, 95, 265004
\refer Cho, J., \& Lazarian, A. 2002, {\it Phys. Rev. Lett.}, 88, 245001
\refer Cho, J., \& Lazarian, A. 2006, {\it Particle Acceleration by MHD Turbulence}, 
\apj 638, 811
\refer Chupp, E. L. 1990, {\it Science}, 250, 229
\refer Cranmer, S. R., \& van Ballegooijen, A. A. 2003, \apj 594, 573
\refer Dennis, B. R., \& Zarro, D. M. 1993, \sphys, 146, 177
\refer Dingus, B. L. et al. 1994, {\it AIP Conf. Proc.} eds. J. M. Ryan \& W. T. Vestrand, 294, 177
\refer Drake, J. F., Paper presented at {\it Krakow Conference on Relativistic Jets}, June  2006
\refer Ellison, D. C., Decourchelle, A, \& Ballet, J. 2005, \aa,  429, 569
\refer Farmer, A., Goldreich, P. 2004, \apj 604, 671
\refer Fisk, L. A. 1978, \apj 224, 1048
\refer Giacalone, J. 2005, \apjl 628, L37
\refer Goldreich, P., \& Sridhar, S. 1995, \apj 438, 763
\refer Goldreich, P., \& Sridhar, S. 1997, \apj 485, 680
\refer Hamilton, R. J. \& Petrosian, V. 1992, \apj 398, 350
\refer Holman, G. D. 1985, \apj {\bf 293}, 584
\refer Holman, G. D. 1996a, {\it BAAS}, 28, 939
\refer Holman, G. D. 1996b,  {\it AIP Conf. Proc.} eds. R. Ramaty et al., 374, 479
\refer Hoshino, M., Paper presented at {\it COSPAR Scientific Assembly}  Beijing, July 2006
\refer Hurford, G. J. et al. 2003, \apj 595, 77L
\refer Ibragimov, I. A. \& Kocharov, G. E. 1977, Proc. 15th Int, Cosmic Ray Conf. (Plovdiv), 11,  340
\refer Iroshnikov, P., 1963, {\it Astron. Zh.}, 40, 742 (English version: 1964,  Sov. Astron., 7, 566)
\refer Jiang, Y. W. et al. 2006,  \apj 638, 1140 
\refer Jones, F. C. 1994, {\it Ap. J. Suppl.}, 90, 561
\refer Jokipii, R. J. 1987, \apj 313, 842
\refer Kraichnan, R., 1965, Phys. Fluids, 8, 1385
\refer Lin, R. P. et al. 1981, \apjl 251, L109
\refer Lithwick, Y., \& Goldreich, P. 2001, \apj 562, 279 
\refer Litvinenko, Y. E. 2003, {\it Solar Phys.}, 212, 379
\refer Liu, S., Petrosian, V., \& Mason, G. 2004, \apjl 613, L81 
\refer Liu, S., Petrosian, V., \& Mason, G. 2006, \apj 634, 462 
\refer Liu, W. et al. 2004, \apj 611, 53L
\refer Liu, W.  2007, from PhD thesis, Stanford University
\refer Luo, Q. \& Melrose, D. 2006; astro-ph/0602295
\refer Kulsrud, R.M. 2005, {\it Plasma Physics for Astrophysics}, Princeton Univ. Press
\refer Mandzhavidze, N., \& Ramaty, R. 1996, {\it BAAS}, 28, 858
\refer Marschh\"{a}user, H. et al. 1994, {\it AIP Conf. Proc.} eds. J. M. Ryan \& W. T. Vestrand 294, 171
\refer Mason, G. M. et al. 1986, \apj 303, 849
\refer Mason, G. M., Dwyer, J. R., \& Mazur, J. E. 2000, \apj 545, 157L
\refer Mason, G. M. et al. 2002, \apj 574, 1039
\refer Masuda, S. et al. 1994, {\it Nature}, 371, 495
\refer Masuda, S. 1994, University of Tokyo, Ph.D. Thesis
\refer Mazur, J. E. et al. 1992, \apj 401, 398
\refer McTiernan, J. M. et al. 1993, \apjl 416, L91
\refer Miller, J. A.\ 2003, {\it COSPAR Colloquia Series} Vol. 13, 387
\refer Miller, J. A.,LaRosa, T.N., \& Moore, R.L. 1996, \apj 445, 464
\refer Miller, J. M. \&  Reames, D. V. 1996, {\it AIP Conf. Proc.} eds. R. Ramaty et al., 374, 450
\refer Miller, J.A.\& Vi\~vas, A. F. 1993, ApJ, 412, 386
\refer Neupert, W.M. 1968, \apj 153, L59
\refer  Park, B. T., Petrosian, V., \& Schwartz, R. A. 1997, \apj 489, 358
\refer Paesold, G., Kallenbach, R. \& Benz, A. O. 2003, ApJ 582, 495 
\refer Petrosian, V. 1973, \apj 186, 291
\refer Petrosian, V., \& Donaghy, T. Q. 1999, \apj 527, 945 
\refer Petrosian, V.,  Donaghy, T. Q., \& McTiernan, J. M. 2002, \apj 569, 459 
\refer Petrosian, V., \& Liu, S. 2004, \apj 610, 550 {\bf PL04}
\refer Petrosian, V., McTiernan, J.M., \& Marschh\"{a}user, H. 1994, \apj 434, 744
\refer Petrosian, V,, Yan, H. \& Lazarian, A. 2006, \apj 644, 603 
\refer Pryadko, J., \& Petrosian, V. 1997, \apj 482, 774 
\refer Pryadko, J., \& Petrosian, V. 1998, \apj 495, 377 
\refer Pryadko, J., \& Petrosian, V. 1999, \apj 515, 873 
\refer Ramaty, R. \& Kozlovsky, B. 1974, ApJ 193, 729
\refer Reames, D. V., Meyer, J. P., \& von Rosenvinge, T. T.\ 1994, ApJS, 90, 649
\refer Reames, D. V. et al. 1997, \apj 483, 515
\refer Sui, L., \& Holman, G. D. 2003, \apjl 596, L251
\refer Sui, L., Holman, G. D., \& Dennis, B. R. 2005, \apj 626, 1102
\refer Swanson, D. G. 1989, {\it Plasma Waves} (New York: Academic Press)
\refer Temerin, M. \&  Roth, I. 1992, ApJ 391, L105
\refer Tsuneta, S. 1985, \apj 290, 353
\refer Veronig, A. M. et al. 2005, \apj 621, 482
\refer Yan, H., \& Lazarian, A. 2002, {\it Phys. Rev. Lett.}, 89, 2881102
\refer Zenitani, S. \& Hoshino, M. 2005, \apj 618, L111
\refer Zhang, T. X. 1995, ApJ, 449, 916
\refer Zhou, Y., \& Matthaeus, W.H. 1990, {\it J. Geophys. Res.} 95, 14881

\end{document}